\documentclass[aps,prd,superscriptaddress,nofootinbib,twocolumn]{revtex4-1}

\usepackage{amsmath}
\usepackage{amssymb}
\usepackage{amsbsy}
\usepackage{amstext}
\usepackage{amsfonts}
\usepackage{mathrsfs}
\usepackage{latexsym}
\usepackage{graphicx}

\begin{document}

\title{Massive gravity and the suppression of anisotropies and gravitational waves in a matter-dominated contracting universe}

\author{Chunshan Lin}
\email{chunshan.lin@yukawa.kyoto-u.ac.jp}
\affiliation{Institute of Theoretical Physics, Faculty of Physics, University of Warsaw, ul.\ Pasteura 5, Warsaw, Poland}
\affiliation{Yukawa Institute for Theoretical Physics, Kyoto University, 606-8502, Kyoto, Japan}

\author{Jerome Quintin}
\email{jquintin@physics.mcgill.ca}
\thanks{Vanier Canada Graduate Scholar}
\affiliation{Department of Physics, McGill University, Montr\'eal, QC, H3A 2T8, Canada}

\author{Robert H.~Brandenberger}
\email{rhb@physics.mcgill.ca}
\affiliation{Department of Physics, McGill University, Montr\'eal, QC, H3A 2T8, Canada}

\begin{abstract}
We consider a modified gravity model with a massive graviton, but which nevertheless only propagates two gravitational degrees of freedom and which is free of ghosts. We show that non-singular bouncing cosmological background solutions can be generated. In addition, the mass term for the graviton prevents anisotropies from blowing up in the contracting phase and also suppresses the spectrum of gravitational waves compared to that of the scalar cosmological perturbations. This addresses two of the main problems of the {\it matter bounce} scenario.
\end{abstract}

\maketitle

\section{Introduction}

The inflationary scenario \cite{inflation} is the current paradigm of very early universe cosmology. It solves
a number of conceptual problems within standard Big Bang cosmology and makes predictions for
the structure in the Universe, which are confirmed to great precision by observations, in particular
the slight red tilt \cite{Mukhanov:1981xt} in the spectrum of scalar cosmological perturbations \cite{Ade:2015lrj}. However, current
realizations of inflation have some conceptual problems (see, e.g., \cite{RHBrev}), in particular
the Trans-Planckian problem for cosmological perturbations \cite{MB}. Hence, it is of interest
to consider possible alternative very early universe scenarios.

In fact, alternatives to cosmological inflation
exist (see, e.g., \cite{RHBrev2} for reviews). In particular, bouncing cosmologies \cite{bouncingreviews} may provide alternatives
to cosmological inflation. The Ekpyrotic scenario \cite{Khoury:2001wf} is one candidate scenario. This
scenario is based on postulating the existence of a new form of matter with an equation of state (EoS)
$p \gg \rho$, where $p$ and $\rho$ are the pressure and energy density, respectively. In this case,
an isotropic phase of contraction is a local attractor in initial condition space \cite{Erickson:2003zm,Garfinkle:2008ei}
(see also \cite{Levy:2015awa} regarding the fine-tuning of the initial conditions), in the same way as for inflationary cosmology
an inflating expanding background is a local attractor in initial condition space \cite{InflAtt}.
However, the spectrum of adiabatic cosmological perturbations has a deep blue spectrum \cite{EkpBlue}, and one
needs to make use of entropy fluctuations to produce a spectrum of nearly scale-invariant curvature
perturbations at late times \cite{NewEkp} (see \cite{Lehners:2008vx} for a review).

The {\it matter bounce} scenario is another alternative to inflation. It has opposite strengths and
weaknesses compared to the Ekpyrotic scenario. On one hand, one does not need to specify any new forms
of matter (except new physics required to obtain a non-singular cosmological bounce). The idea is that the
universe starts in a homogeneous and isotropic contracting phase with the same matter content of
the current expanding universe, i.e.~with cold matter, radiation, and possibly a very small
cosmological constant, required to explain the currently observed dark energy component.
Then, it can be shown \cite{Finelli:2001sr,Wands:1998yp} that adiabatic fluctuations with comoving wavelengths
which originate in their quantum vacuum state and exit the
Hubble radius during the matter-dominated phase of contraction acquire an almost scale-invariant
spectrum at late times in the contracting phase. In fact, the presence of the dark energy component
leads to a slight red tilt of the scalar spectrum \cite{LCDMbounce}, in agreement with the observed spectrum.
On the other hand, the homogeneous and isotropic contracting trajectory is not an attractor in initial
condition space. In fact, the energy density in anisotropies grows faster than the energy density
in the matter components, leading to an instability of the model \cite{Cai:2013vm}, known as the
Belinsky-Khalatnikov-Lifshitz (BKL) instability \cite{Belinsky:1970ew}. This problem is usually evaded
with the inclusion of either higher-order curvature terms in the gravity action \cite{Middleton:2008rh},
a phase of Ekpyrotic contraction (see the models studied in \cite{matterEkpyrotic,Cai:2013vm}),
or another source with ultra-stiff EoS \cite{Bozza:2009jx}.
However, these resolutions are often fine-tuned \cite{Levy:2016xcl,Bozza:2009jx}
or simply not robust to all types of anisotropies \cite{Barrow:2015wfa}.

A second problem for matter bounce scenarios is that the scalar cosmological perturbations and
gravitational waves grow at the same rate on super-Hubble scales since the squeezing factors
in their mode equations of motion (EOMs) are the same (see, e.g., \cite{Mukhanov:1990me,Brandenberger:2003vk}
for reviews of the theory of cosmological perturbations). Because of that,
the scalar and tensor power spectra have the same amplitude at the end of the contracting phase,
i.e., the tensor-to-scalar ratio $r$ is of order unity (see, e.g., \cite{Cai:2008qw,Cai:2014xxa,Quintin:2015rta,Li:2016xjb}).
In addition, the amplitude of non-Gaussianities (characterized by the quantity $f_\mathrm{NL}$) is
of order unity \cite{Cai:2009fn} (see, however, \cite{GLP}). Although it is possible to construct single field bounce models which boost
the scalar fluctuations relative to the gravitational waves, these mechanisms typically also
boost the non-Gaussianities to a level which is in contradiction with the current limits,
which led to the conjecture of a no-go theorem for the class of single field matter bounce
models \cite{Quintin:2015rta,Li:2016xjb}. We call this the {\it large $r$ problem}.

In this paper, we suggest a possible solution to both of these problems of the matter bounce scenario in
the context of a modified gravity model with a massive graviton.
The idea of modifying general relativity so that the graviton acquires a non-trivial mass has been extensively studied
(see, e.g., the review \cite{deRham:2014zqa} and also \cite{massivecosmo} in the context of cosmology),
especially as an attempt to explaining the accelerated expansion of the universe (see, e.g., \cite{GLM}).
Massive gravity has also been studied in the context of very early universe cosmology, e.g.,
during inflation, in which case the propagation of gravitational waves would be affected by the non-trivial
mass of the graviton \cite{Gumrukcuoglu:2012wt,Lin:2015nda,Kuroyanagi:2017kfx} (see also \cite{Domenech:2017kno}).
Using this setup in the context of matter bounce cosmology, we find that the fluctuation equation for the gravitational waves
has a mass term which prevents the squeezing of the
modes on super-Hubble scales and hence solves the large $r$ problem. As it turns out,
the massive graviton also leads to a mass term in the EOM for the anisotropy
parameter, and hence provides a natural isotropization mechanism in the contracting phase.

In the following, we first introduce the modified gravity theory we will be using. A Hamiltonian
analysis shows that the theory is free of ghosts at the fully non-perturbative level and contains only
two propagating gravitational modes. In section \ref{sec:backgroundfull} we study the background evolution
and show that the functions appearing in
the gravitational action can be chosen such that a non-singular cosmological bounce results.
In section \ref{sec:pertanalysis} we study the evolution of cosmological fluctuations in the theory, discussing
scalar, vector, and tensor modes. In particular, we show how the mass term due to the non-trivial graviton mass
arises in the gravitational wave EOMs.
In section \ref{sec:anisotropies} we show how a mass term also arises in the EOMs governing the anisotropies.
Section \ref{sec:solmatter} presents explicit solutions for the anisotropies and for the graviton spectrum
in a matter phase of contraction. We discuss both the limits of large and small graviton
masses and connect the results with observations. We summarize and discuss our results in section \ref{sec:conclusions}.
 
A word concerning notation: we work in natural units ($c = \hbar = 1$) with
the reduced Planck mass and time defined by $M_\mathrm{Pl}\equiv t_\mathrm{Pl}^{-1}\equiv 1/\sqrt{8\pi G_\mathrm{N}}$, and we assume for
computational simplicity that the universe is spatially flat.

\section{The modified gravity theory}

\subsection{Setup}

Our goal is to construct a modified theory of gravity that allows for a non-trivial graviton mass, while being as close as possible to Einstein
gravity. In order to achieve this, we work with the Arnowitt-Deser-Misner (ADM) \cite{Arnowitt:1962hi} decomposition of the
four-dimensional metric $g_{\mu\nu}$,
\begin{equation}
 g_{\mu\nu}\mathrm{d}x^\mu\mathrm{d}x^\nu=-N^2\mathrm{d}t^2+h_{ij}\left(\mathrm{d}x^i+N^i\mathrm{d}t\right)\left(\mathrm{d}x^j+N^j\mathrm{d}t\right)\,,
\end{equation}
where $N(t,\mathbf{x})$ and $N^i(t,\mathbf{x})$ are the lapse and shift functions, respectively,
and $h_{ij}$ is the three-dimensional induced metric tensor on the spatial hypersurface.
In our theory, the graviton mass arises\footnote{The following setup to generate a non-trivial graviton mass
is a generalization of the Lorentz-violating massive gravity theory of \cite{Dubovskyrefs}.}
from the non-trivial vacuum expectation value of four St\"{u}ckelberg
scalar fields,
\begin{equation}
 \varphi^0(t,\mathbf{x})=M^2f(t)\,, \qquad \varphi^a(t,\mathbf{x})=M^2x^i\delta_i{}^a\,,
\end{equation}
where $a\in\{1,2,3\}$ and the St\"{u}ckelberg
scalar fields have mass dimension ($f(t)$ has dimension $[-1]$ and $M$ is a mass scale). In the scalar field configuration, the following internal
symmetries are imposed \cite{Dubovskyrefs},
\begin{equation}
\label{eq:sym1}
 \varphi^a\to\Lambda^a{}_b\varphi^b\,, \qquad \varphi^a\to\varphi^a+\Xi^a(\varphi^0)\,,
\end{equation}
where $\Lambda^a{}_b$ is the $SO(3)$ rotation operator, and $\Xi^a(\varphi^0)$ are three generic functions of their argument.
The internal symmetry between the space-like St\"{u}ckelberg scalar fields $\varphi^a$ and the time-like
St\"{u}ckelberg scalar field $\varphi^0$, i.e.\ $\varphi^a\to\varphi^a+\Xi^a(\varphi^0)$, is crucial to eliminate the vector
modes in the gravity sector, as we will see later in section \ref{sec:vecpert}. This symmetry
projects out all temporal derivative terms of $\varphi^a$, and therefore, one finds at the end that these scalar fields are actually
non-dynamical. A more rigorous proof will also be given with the Hamiltonian analysis in the next
sub-section (see section \ref{sec:Hamiltoniananalysis}).

At the first derivative level, we have the following quantity that respects the symmetries of equation \eqref{eq:sym1}
(see \cite{Dubovskyrefs}),
\begin{equation}\label{Zij}
 Z^{ab}=g^{\mu\nu}\partial_{\mu}\varphi^a\partial_{\nu}\varphi^b
 -\frac{(g^{\mu\nu}\partial_{\mu}\varphi^0\partial_{\nu}\varphi^a)
 (g^{\kappa\xi}\partial_{\kappa}\varphi^0\partial_{\xi}\varphi^b)}{X}\,,
\end{equation}
where $X\equiv g^{\mu\nu}\partial_\mu\varphi^0\partial_\nu\varphi^0$
represents the kinetic term of the time-like St\"{u}ckelberg scalar field $\varphi^0$.
We note that in unitary gauge, one has $Z^{ab}=M^4h^{ij}\delta_i{}^a\delta_j{}^b$.
Then, the graviton mass term can be written as a generic function of $Z^{ab}$. 
However, it is useful to first define the following traceless tensor \cite{Lin:2015cqa,Lin:2015nda},
\begin{equation}
 \bar{\delta}Z^{ab}\equiv\frac{Z^{ab}}{Z}-3\frac{Z^a{}_cZ^{cb}}{Z^2}\,,
\end{equation}
where the internal indices are raised or lowered with $\delta_{ab}$, i.e., $Z\equiv Z^{ab}\delta_{ab}$,
$Z^a{}_{c}\equiv Z^{ad}\delta_{dc}$, and $Z_{ab}\equiv Z^{cd}\delta_{ca}\delta_{db}$.
The graviton mass is then written in terms of the contraction of this traceless tensor,
\begin{equation}
\label{eq:Lmassterm}
 \mathcal{L}_\mathrm{mass}\sim M_\mathrm{Pl}^2m_g^2\bar{\delta}Z^{ab}\bar{\delta}Z_{ab}\,,
\end{equation}
which is a term that breaks space-time diffeomorphism invariance in unitary gauge, and $m_g=m_g(t,Z^{ab},\delta_{ab})$
is a scalar function of its arguments in the internal space.
The resulting action of our minimally-modified effective field theory (EFT) of gravity reads
\begin{align}
 S=&\int\mathrm{d}^3\mathbf{x}\,\mathrm{d}t~\sqrt{h}\left[N\left(\frac{M_\mathrm{Pl}^2}{2}{}^{(4)}\!R-\Lambda_1(t)\right)\right. \nonumber \\
 &\left.-\frac{9}{8}M_\mathrm{Pl}^2m_g^2\bar{\delta}Z^{ab}\bar{\delta}Z_{ab}-\Lambda_2(t)\right]\,, \label{mmg1}
\end{align}
where $\Lambda_1(t)$ and $\Lambda_2(t)$ are some functions of time, which will be fixed later by the EOMs.
The four-dimensional Ricci scalar is denoted by ${}^{(4)}\!R$, and according to the ADM formalism, it can be decomposed as
\begin{equation}
 {}^{(4)}\!R=K_{ij}K^{ij}-K^2+{}^{(3)}\!R+\mathrm{total\ derivatives}\,,
\end{equation}
where $K_{ij}$ is the extrinsic curvature tensor, $K\equiv h^{ij}K_{ij}$ is its trace,
and ${}^{(3)}\!R$ is the three-dimensional Ricci scalar on the spatial hypersurface.
Accordingly, due to the broken temporal diffeomorphism invariance,
the action could actually be written in even more generality as
\begin{align}
 S=&\int\mathrm{d}^3\mathbf{x}\,\mathrm{d}t~\sqrt{h}\bigg[N\bigg(\frac{M_\mathrm{Pl}^2}{2}\Big[c_1(t)\left(K_{ij}K^{ij}-K^2\right) \nonumber \\
 &+c_2(t){}^{(3)}\!R\Big]-\Lambda_1(t)\bigg)-\frac{9}{8}M_\mathrm{Pl}^2m_g^2\bar{\delta}Z^{ab}\bar{\delta}Z_{ab} \nonumber \\
 &-\Lambda_2(t)\bigg]\,,
\label{eq:actionmoregeneral}
\end{align}
of which phase space has even dimension, hence it is free from pathological inconsistencies \cite{Lin:2017oow}.
With the above action, we would find that the propagation speed of gravitational waves is given by
$c_g(t)=\sqrt{c_2(t)/c_1(t)}$. However, in what follows, we will often set $c_1=c_2=1$ so that\footnote{Thus, the model is
consistent with the constraints coming from
the joint observations of gravitational waves and electromagnetic
signals from GW170817 \cite{GBM:2017lvd}.} $c_g=1$.
The $m_g^2$ term is the graviton mass term, and it does not contribute to the background evolution.
Its non-trivial contribution only appears in the equations for perturbations.
In particular, this theory is a generalization of one of the theories discussed in \cite{Lin:2017oow}
(see also \cite{Lin:2017utd}).

In the action of equation \eqref{mmg1},
we note that $\Lambda_1$ is akin to a cosmological constant as in $\Lambda$CDM cosmology.
However, we allow it to be time-dependent, which can be done in a consistent manner as long as the appropriate constraint equation is satisfied
as we will see in section \ref{sec:recovdiff}. This is similar to the constraint equation that follows from allowing the cosmological
constant, Newton's constant, and the speed of light to actually be non-constant (see, e.g., \cite{varyingconstants}).
Here, the key is that we allow for another cosmological constant-like function, $\Lambda_2$, which is independent of the lapse function
in the EFT and which can also be time dependent.
This type of function can appear in different modified theories of gravity, for instance in Cuscuton cosmology \cite{cuscutoncosmo}.

Note that we are working in unitary gauge. As always, rather than starting from an EFT,
it is possible to recover the four-dimensional general covariance
by introducing a scalar field, e.g.~$\Lambda_1(t)\to \Lambda_1(\varphi^{0}(t,\textbf{x}))$ and so on.
From the point of view of such a covariant description of the same theory, the action of equation \eqref{mmg1} written
in terms of $N$, $N^i$, $h_{ij}$, and without the scalar field, is nothing but the action in the so-called unitary gauge,
in which the time coordinate is chosen to
agree with a fixed monotonic function of the scalar field. 

As we will see later, this theory is ghost free and able to realize a non-singular bounce, and at the same time, it yields a mass correction
to gravitational waves and anisotropies. It also has only two degrees of freedom (DOFs) in the gravity sector as we will now demonstrate.

\subsection{Hamiltonian analysis}\label{sec:Hamiltoniananalysis}

The conjugate momenta of the theory are given by
\begin{align}
 \pi^{ij}&\equiv\frac{\partial\mathcal{L}}{\partial\dot{h}_{ij}}=\frac{M_\mathrm{Pl}^2}{2}\sqrt{h}\left(K^{ij}-Kh^{ij}\right)\,, \\
 \pi_N&\equiv\frac{\partial\mathcal{L}}{\partial\dot{N}}=0\,, \\
 \pi_i&\equiv\frac{\partial\mathcal{L}}{\partial\dot{N}^{i}}=0\,,
\end{align}
where the Lagrangian density $\mathcal{L}$ can be read off from the action of equation \eqref{mmg1},
and a dot denotes a derivative with respect to physical time $t$.
The Hamiltonian then reads
\begin{align}
 \mathsf{H}=&~\int\mathrm{d}^3\mathbf{x}~\left(\pi^{ij}\dot{h}_{ij}-\mathcal{L}+\lambda_N\pi_N+\lambda^i\pi_i\right) \\
 =&~\int\mathrm{d}^3\mathbf{x}~\Big[N\mathcal{C}+N^i\mathcal{H}_i+\lambda_N\pi_N+\lambda^i\pi_i \nonumber \\
 &-\sqrt{h}\,G(h^{ij},t)\Big]\,.
\end{align}
In the above, $\lambda_N$ and the $\lambda^i$'s are Lagrange multipliers, and so, we have
\begin{equation}
\label{eq:primcons}
 \pi_N\approx 0 \qquad \mathrm{and} \qquad \pi_i\approx 0
\end{equation}
as primary constraints. Also, we defined
\begin{equation}
 G(h^{ij},t)\equiv -\frac{9}{8}M_\mathrm{Pl}^2m_g^2\bar{\delta}Z^{ab}\bar{\delta}Z_{ab}-\Lambda_2(t)\,,
\end{equation}
which represents the lapse-independent terms in the action, and
\begin{align}
 &\mathcal{C}\equiv\frac{2}{M_\mathrm{Pl}^2\sqrt{h}}\left(\pi^{ij}\pi_{ij}-\frac{1}{2}\pi^2\right)
 -\frac{M_\mathrm{Pl}^2}{2}\sqrt{h}\,{}^{(3)}\!R+\sqrt{h}\,\Lambda_1(t)\,, \\
 &\mathcal{H}_i\equiv -2\sqrt{h}D_j\left(\frac{\pi_i^{~j}}{\sqrt{h}}\right)\,.
\end{align}
In the above, $D_i$ is the covariant derivative on the three-dimensional spatial hypersurface,
and the indices of $\pi^{ij}$ are lowered or raised with the three-dimensional metric tensor,
i.e., $\pi\equiv h_{ij}\pi^{ij}$, $\pi_i{}^j\equiv h_{il}\pi^{lj}$, etc.
The consistency conditions of the four primary constraints in equation \eqref{eq:primcons} give us
\begin{align}
 \frac{\mathrm{d}\pi_N}{\mathrm{d}t}&=\{\pi_N,\mathsf{H}\}=-\mathcal{C}\approx 0\,, \\
 \frac{\mathrm{d}\pi_i}{\mathrm{d}t}&=\{\pi_i,\mathsf{H}\}=-\mathcal{H}_i\approx 0\,,
\end{align}
which represent the Hamiltonian and momentum constraints, respectively, and they are the secondary constraints.
The consistency condition of the Hamiltonian constraint gives us the following tertiary constraint,
\begin{align}\label{hconsis}
 \frac{\mathrm{d}\mathcal{C}}{\mathrm{d}t}=&~\frac{\partial \mathcal{C}}{\partial t}+\{\mathcal{C},\mathsf{H}\}
 =\sqrt{h}\,\frac{\partial\Lambda_1}{\partial t} \nonumber \\
 &+\frac{4}{M_\mathrm{Pl}^2\sqrt{h}}\left(\pi_{ij}-\frac{1}{2}\pi h_{ij}\right)
 \frac{\partial}{\partial h_{ij}}\left(\sqrt{h}\,G(h^{ij},t)\right) \nonumber \\
\equiv &~\mathcal{C}_3\approx ~0\,.
\end{align}
On the other hand, the consistency conditions of the momentum constraints give us the following three tertiary constraints,
\begin{equation}
 \frac{\mathrm{d}\mathcal{H}_i}{\mathrm{d}t}=\frac{\partial\mathcal{H}_i}{\partial t}+\{\mathcal{H}_i,\mathsf{H}\}
 =\{\mathcal{H}_i,\mathsf{H}\}\equiv\mathcal{H}_{3,i}\approx 0\,.
\end{equation}

We can explicitly check that the one Hamiltonian constraint, three momentum constraints, as well as the corresponding four
tertiary constraints are all of second class. On the other hand, the four primary constraints in equation \eqref{eq:primcons}
are of first class. Therefore, since each first class constraint eliminates two phase space degrees of freedom and each
second class constraint eliminates one such degree of freedom, the total number of configuration space DOFs is
\begin{align}
 \#\mathrm{DOFs}&=\frac{1}{2}\left[2\times 10-2\times\left(1+3\right)-\left(1+3+1+3\right)\right] \nonumber \\
 &=2\,.
\end{align}
Consequently, the graviton has only two tensor polarizations, and there is no scalar or vector gravitons in the theory.
This is in agreement with the general analysis of \cite{Comelli:2014xga}.
This will also be explicit from the analysis of metric perturbations in section \ref{sec:pertanalysis}.
Another similar example in which the graviton has only two polarizations can be found in \cite{DeFelice:2015hla}.

\subsection{Recovering the space-time diffeomorphism invariance}\label{sec:recovdiff}

The space-time diffeomorphism symmetry is broken in the action of equation \eqref{mmg1}.
To recover the space-time diffeomorphism invariance, $Z^{ab}$ should be rewritten in terms of scalar fields, i.e.~as in equation \eqref{Zij},
and let us also write
\begin{align}
 &\Lambda_1(t)&\to&~ V(\varphi^0)\,, \\
 &\Lambda_2(t)&\to&~ M^2N\sqrt{-X}\,, \\
 &m_g^2(t,Z^{ab},\delta_{ab})&\to&~ N\sqrt{-X}F(\varphi^0,Z^{ab},\delta_{ab})\,,
\label{eq:mgtcov}
\end{align}
where $\varphi^0(t,\textbf{x})=M^2f(t)+\delta\varphi^0(t,\textbf{x})$, $F$ is a generic dimensionless function of its arguments,
and we assume that $f(t)$ is a monotonic function of time
(otherwise there may be some ambiguity in the unitary gauge).
Then, in a Friedmann-Lema\^{i}tre-Robertson-Walker (FLRW) background,
\begin{equation}
 g_{\mu\nu}\mathrm{d}x^\mu\mathrm{d}x^\nu=-\mathrm{d}t^2+a(t)^2\delta_{ij}\mathrm{d}x^i\mathrm{d}x^j\,,
\end{equation}
and assuming that $\dot{\varphi}^0>0$ (thus $X=-N^{-2}(\dot{\varphi}^0)^2$ and $\Lambda_2=M^2\dot{\varphi}^0$),
the background action of the time-like St\"{u}ckelberg scalar field reads
\begin{equation}
 S_{\varphi^0}=-\int\mathrm{d}^3\mathbf{x}\,\mathrm{d}t~a^3\left[V(\varphi^0)+M^2\dot{\varphi}^0\right]\,.
\end{equation}
Taking the variation of the above action with respect to $\varphi^0$, we get 
\begin{equation}
\label{eq:constrainteq}
 3H=\frac{V'(\varphi^0)}{M^2}=\frac{\dot{V}}{M^2\dot{\varphi}^0}=\frac{\dot{\Lambda}_1}{\Lambda_2}\,,
\end{equation}
where we use the chain rule in the second equality,
and $H\equiv\dot{a}/a$ represents the Hubble parameter.
We note that equation \eqref{eq:constrainteq} is a constraint equation that determines the allowed functional form of $\Lambda_1$ and $\Lambda_2$
in the EFT for a given FLRW background.

\section{Nonsingular bouncing cosmology}\label{sec:backgroundfull}

\subsection{Background evolution}\label{sec:background}

We are interested in studying the matter bounce scenario with the action of equation \eqref{mmg1}.
We first study the background evolution in this context, which should consist of a least a matter-dominated
contracting phase and a non-singular bouncing phase. The bouncing phase serves as a transition from contraction to expansion,
and we should recover standard Big Bang cosmology in the expanding phase.

In an FLRW background, the consistency condition of the Hamiltonian constraint, i.e.\ equation \eqref{hconsis},
reduces to\footnote{The generalization with the action of equation \eqref{eq:actionmoregeneral} is
\begin{equation}
 \dot\Lambda_1+6M_\mathrm{Pl}^2H^2\dot c_1-3H\Lambda_2=0\,.
\end{equation}}
\begin{equation}
\label{hcon2}
 \dot{\Lambda}_1=3H\Lambda_2.
\end{equation}
This equation is actually the EOM of the non-dynamical St\"{u}ckelberg scalar field $\varphi^0$ if we want to recover
the general covariance of the theory.
Indeed, we notice that it is the same as equation \eqref{eq:constrainteq} that one finds when recovering the space-time diffeomorphism
invariance of the action.

To represent the matter sector of the theory, one can introduce a canonical scalar field $\phi$ minimally coupled to gravity, 
\begin{equation}
 S_\mathrm{matter}=\int\mathrm{d}^4x~\sqrt{-g}\left(-\frac{1}{2}g^{\mu\nu}\partial_\mu\phi\partial_\nu\phi-U(\phi)\right)\,.
\label{eq:Smatterfield}
\end{equation}
The Einstein equations then read\footnote{Again, for the sake of completeness, these can be generalized to
\begin{align}
 3M_\mathrm{Pl}^2c_1H^2&=\frac{1}{2}\dot{\phi}^2+U(\phi)+\Lambda_1\,, \\
 M_\mathrm{Pl}^2c_1\dot{H}&=-\frac{1}{2}\dot{\phi}^2-M_\mathrm{Pl}^2H\dot c_1+\frac{\Lambda_2}{2}\,.
\end{align}}
\begin{align}
\label{eq:Friedmann1}
 3M_\mathrm{Pl}^2H^2&=\frac{1}{2}\dot{\phi}^2+U(\phi)+\Lambda_1\,, \\
\label{eq:Friedmann2}
 M_\mathrm{Pl}^2\dot{H}&=-\frac{1}{2}\dot{\phi}^2+\frac{\Lambda_2}{2}\,,
\end{align}
and the variation of equation \eqref{eq:Smatterfield} with respect to $\phi$ yields
\begin{equation}
\label{eq:KGphi}
 \ddot{\phi}+3H\dot{\phi}+U'(\phi)=0\,.
\end{equation}
One can check that the two Friedmann equations \eqref{eq:Friedmann1} and \eqref{eq:Friedmann2}
are consistent with each other, provided the constraint equation \eqref{hcon2} and the EOM of matter [equation \eqref{eq:KGphi}] are satisfied.
We can input the time dependence of $\Lambda_2(t)$ in the first place, and then, $\Lambda_1(t)$ evolves according to equation \eqref{hcon2}.
In order to transition from a contracting universe ($H<0$) to an expanding universe ($H>0$) through a non-singular bounce,
one needs to violate the Null Energy Condition (NEC), which in this case is equivalent to requiring the condition $\dot{H}>0$.
Thus, from equation \eqref{eq:Friedmann2}, the condition is simply $\Lambda_2>\dot{\phi}^2$ during the non-singular bouncing phase.

Alternatively, one could describe the background evolution by simply introducing a general fluid with energy-momentum tensor
$T_{\mu\nu}\equiv (2/\sqrt{-g})\delta S_\mathrm{matter}/\delta g^{\mu\nu}$ such that
at the background level $T_\mu{}^\nu=\mathrm{diag}(-\rho(t),p(t)\delta_i{}^j)$. Then,
the EOMs are (with $c_1\equiv 1$)
\begin{align}
\label{eq:Friedmann12}
 &3M_\mathrm{Pl}^2H^2=\rho+\Lambda_1\,, \\
\label{eq:Friedmann22}
 &2M_\mathrm{Pl}^2\dot H=-(\rho+p)+\Lambda_2\,,
\end{align}
together with the conservation equation
\begin{equation}
 \dot\rho+3H(\rho+p)=0\,.
\end{equation}
Then, in order to consider the matter bounce scenario, let us assume that the matter content has an approximately vanishing pressure,
i.e.\ $p\approx 0$, in which case the conservation equation immediately implies that $\rho\propto a^{-3}$.
Also, this means that the NEC can be violated if $\Lambda_2>\rho$.

Let us now construct a specific example that reproduces the desired background evolution.
Let us take the following ansatz,
\begin{equation}
\label{eq:Lambda2ansatz}
 \Lambda_2(t)=\Lambda_{2,0}\exp\left(-\frac{(t-t_B)^2}{\sigma^2}\right)\,,
\end{equation}
where $\Lambda_{2,0}$, $\sigma$, and $t_B$ are free constant parameters at this point.
It is clear that in the limit where $|t-t_B|\gg\sigma$, one finds that $\Lambda_2\simeq 0$,
and thus, $\dot\Lambda_1=3H\Lambda_2\simeq 0$, hence $\Lambda_1\simeq\mathrm{constant}$ (let us call
this constant $\Lambda_{1,0}$).
In the limit where $|t-t_B|\ll\sigma$, one finds that $\Lambda_2\simeq\Lambda_{2,0}$.
This suggests the existence of three regimes.

First, as $t\rightarrow -\infty$, we have $a\rightarrow\infty$ and $\rho\propto a^{-3}\rightarrow 0$,
so $\Lambda_1\rightarrow\Lambda_{1,0}\gg \rho$, and thus in this limit equations \eqref{eq:Friedmann12} and \eqref{eq:Friedmann22} become
\begin{equation}
 H\rightarrow-\frac{1}{M_\mathrm{Pl}}\sqrt{\frac{\Lambda_{1,0}}{3}}\,,\qquad \dot H\rightarrow 0\,.
\end{equation}
This is the cosmological constant-dominated contracting phase.

\begin{figure*}
 \centering
 \includegraphics[scale=0.80]{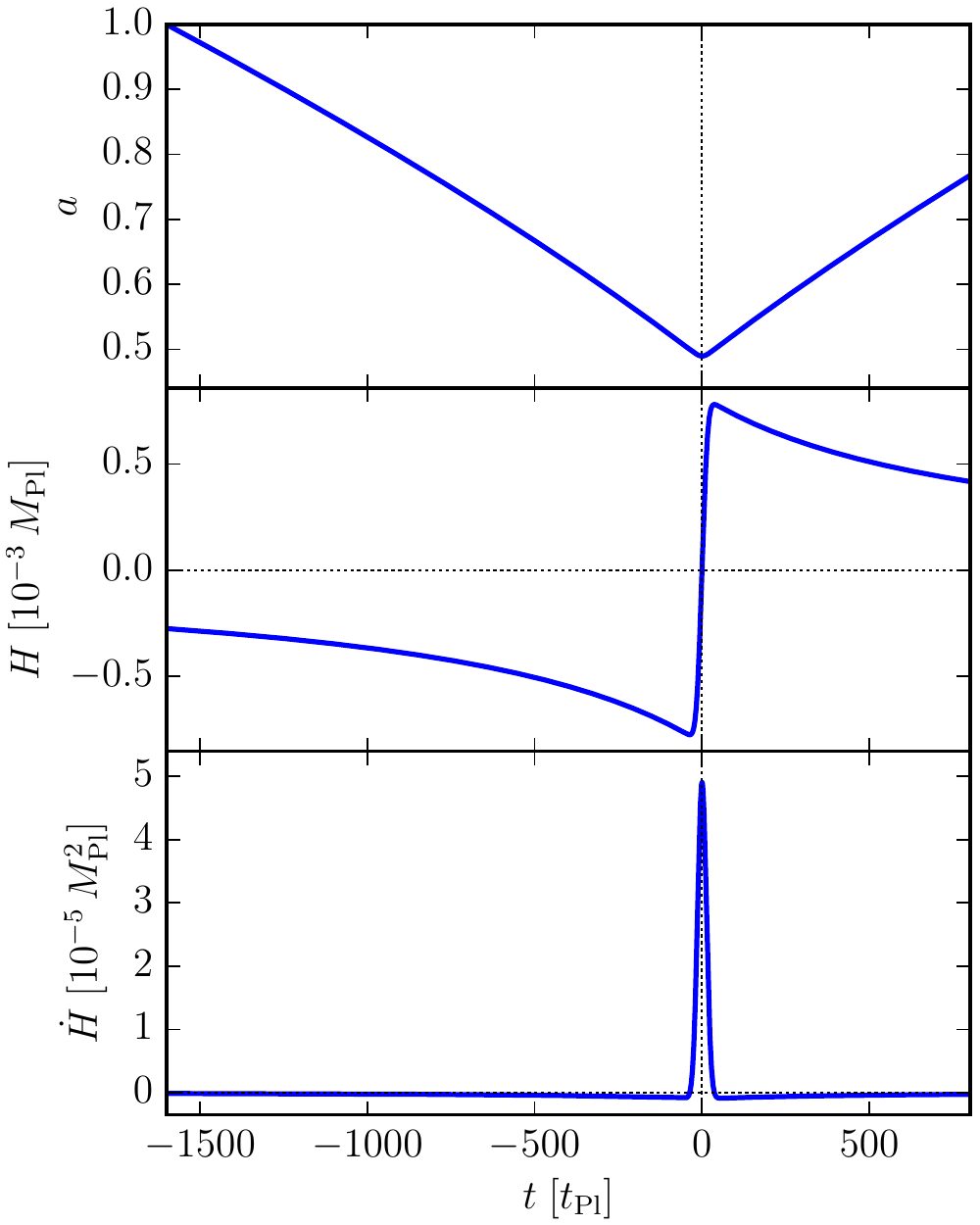}
 \hspace*{1cm}
 \includegraphics[scale=0.80]{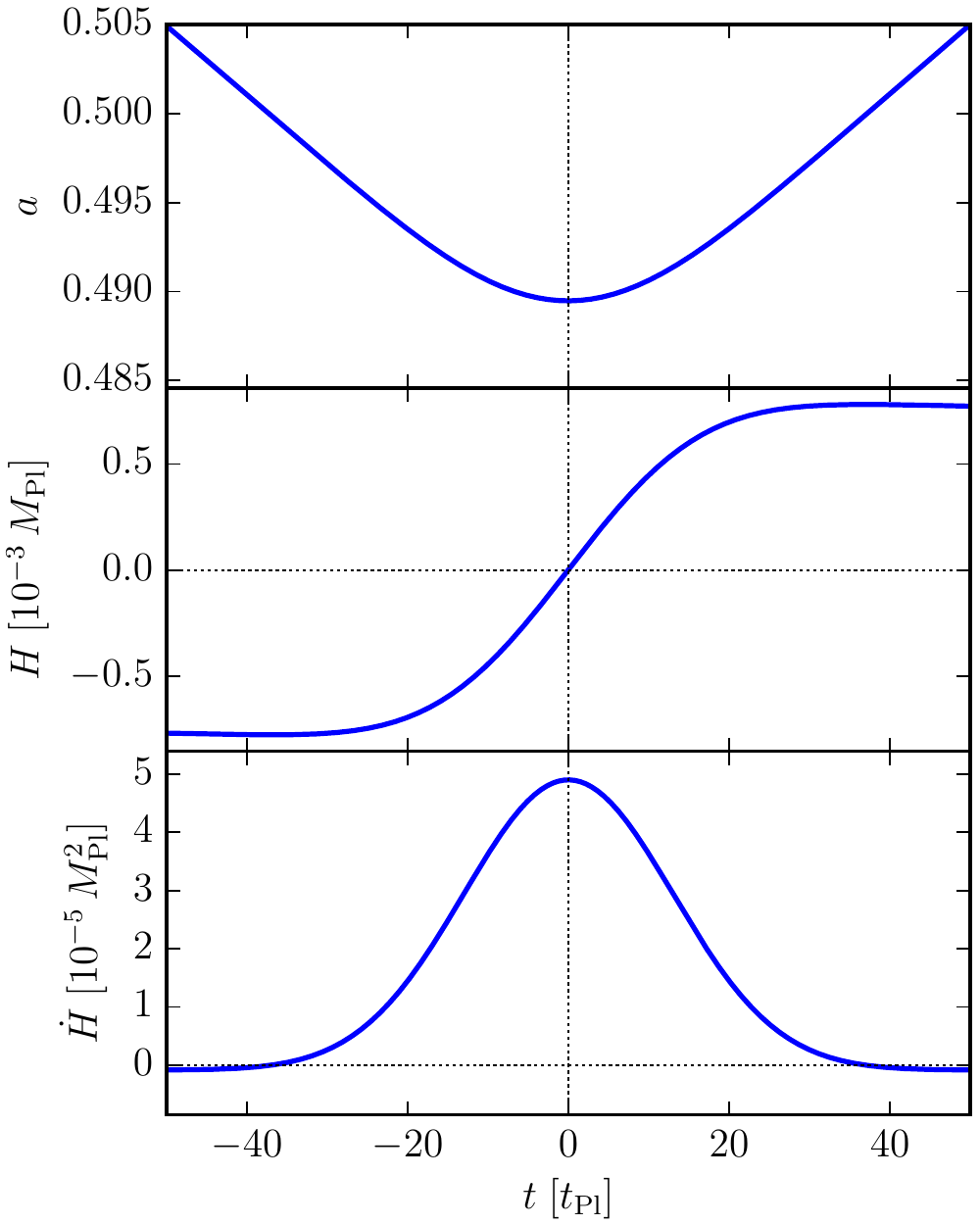}
 \caption{Background evolution for the matter bounce model presented in section \ref{sec:background}.
 In particular, $\Lambda_2(t)$ is taken according to equation \eqref{eq:Lambda2ansatz} with
 $\Lambda_{2,0}=1.0\times 10^{-4}\,M_\mathrm{Pl}^4$ and $\sigma=18.441\,t_\mathrm{Pl}$.
 These numbers are chosen such that $|H(t_{B-})|=8.0\times 10^{-4}\,M_\mathrm{Pl}$,
 where $t_{B-}$ represents the time when the NEC starts being violated (so $\dot{H}(t_{B-})=0$).
 In other words, $H(t_{B-})$ is the value of the Hubble parameter at the end of the contracting phase,
 just before the bouncing phase, so it is the maximum of the absolute value of the Hubble parameter.
 Note that we set the bounce point at $t_B=0$.
 Then, one can solve equations \eqref{eq:Friedmann12} and \eqref{eq:Friedmann22} together with equation \eqref{hcon2} with $p=0$
 and $\rho=\rho_0(a_0/a)^3$.
 The initial conditions are taken at the time $t_\mathrm{ini}=-1.6\times 10^3\,t_\mathrm{Pl}$ as follows: $a(t_\mathrm{ini})=1$,
 $H(t_\mathrm{ini})=-2.758\times 10^{-4}\,M_\mathrm{Pl}$, and $\rho(t_\mathrm{ini})=2.282\times 10^{-7}\,M_\mathrm{Pl}^4$.
 The plots above show the evolution of the scale factor $a$ as well as the Hubble parameter $H=\dot{a}/a$ and
 its time derivative $\dot H$. The plots on the left show the evolution from the time at which the initial conditions are set
 until slightly after the bouncing phase. The plots on the right show a blowup of the same evolution
 restricted to the bouncing phase.}
\label{fig:background}
\end{figure*}

Second, there is a regime during which $|t-t_B|\gg\sigma$, so $\Lambda_1\simeq\Lambda_{1,0}$
and $\Lambda_2\simeq 0$, but where we still have $\rho\gg\Lambda_1$. If we write
$\rho(t)=\rho_0[a_0/a(t)]^3$, then there is a time $t_\mathrm{eq}$ when
$\rho(t_\mathrm{eq})=\Lambda_{1,0}$, and thus, $\rho(t)\gg\Lambda_1$
for $\sigma\ll|t-t_B|\ll t_\mathrm{eq}$.
In that case, $3M_\mathrm{Pl}^2H^2\simeq\rho\propto a^{-3}$,
and so, $a(t)\propto(-t)^{2/3}$ and $H\sim 2/(3t)$.
This is the matter-dominated contracting phase.

Third, in the regime when $|t-t_B|\ll\sigma$, we have $\Lambda_2\simeq\Lambda_{2,0}$, and so,
if $\Lambda_{2,0}\gg\rho(t)$ in that time interval, we have $2M_\mathrm{Pl}^2\dot H\simeq\Lambda_{2,0}$.
Consequently,
\begin{equation}
\label{eq:Hbounce}
 H(t)\simeq\frac{\Lambda_{2,0}}{2M_\mathrm{Pl}^2}(t-t_B)\,,
\end{equation}
where we set the integration constant such that $H(t_B)=0$, i.e., $t=t_B$ is the bounce point where the transition
from $H<0$ to $H>0$ occurs.
Therefore, this is the non-singular bouncing phase.
In that case, $\dot\Lambda_1=3H\Lambda_2\simeq 3\Lambda_{2,0}^2(t-t_B)/(2M_\mathrm{Pl}^2)$,
which implies
\begin{equation}
\label{eq:Lambda1bounce}
 \Lambda_1(t)\simeq\frac{3\Lambda_{2,0}^2}{4M_\mathrm{Pl}^2}(t-t_B)^2+C\,,
\end{equation}
where $C$ is another integration constant.
Requiring $H(t_B)=0$ implies $\rho(t_B)+\Lambda_1(t_B)=0$,
hence $C=-\rho(t_B)=-\rho_0(a_0/a_B)^3$ with $a_B\equiv a(t_B)$.

\begin{figure}
 \centering
 \includegraphics[scale=0.80]{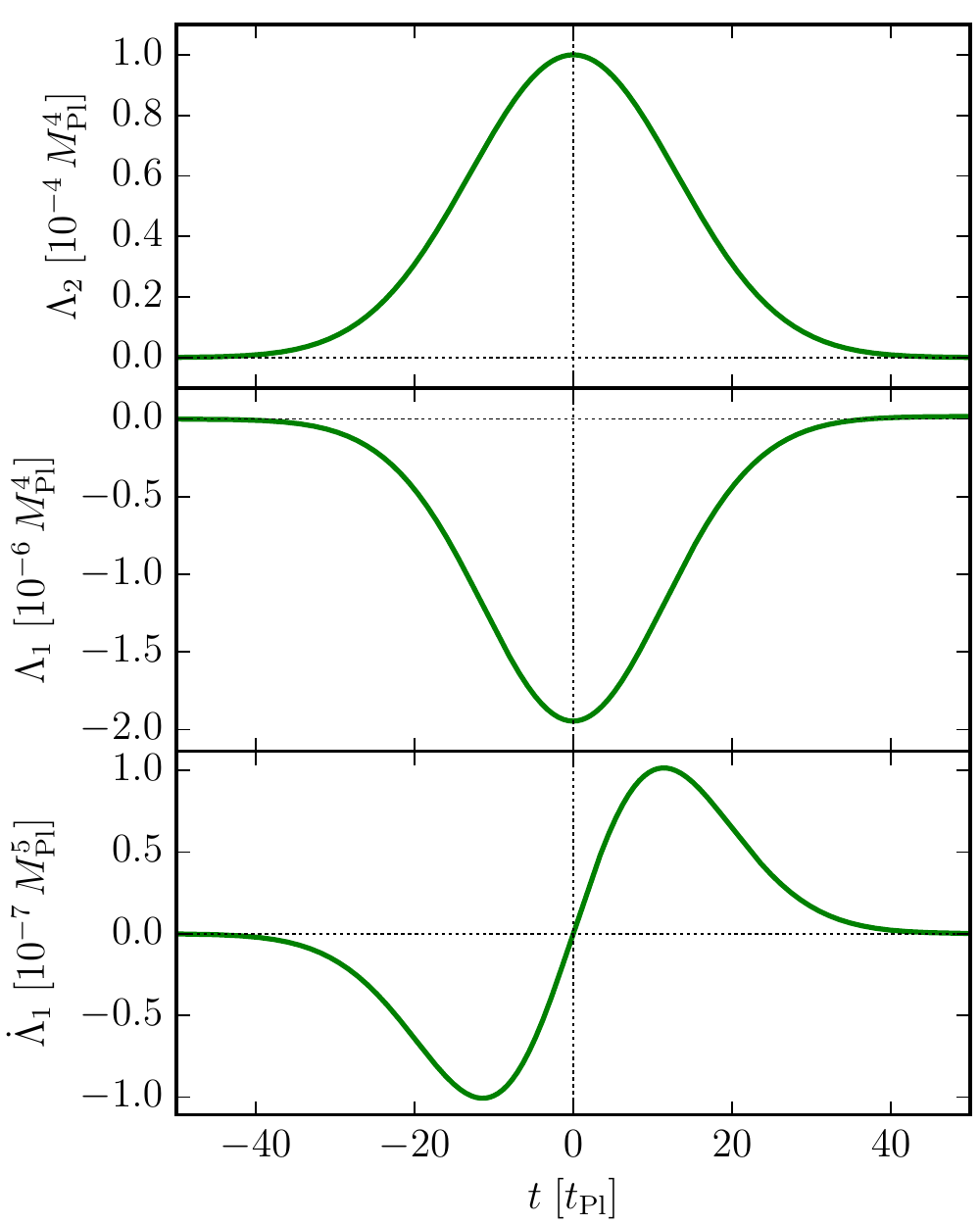}
 \caption{Plots of the lapse-independent function in the EFT $\Lambda_2(t)$, the cosmological constant-like function $\Lambda_1(t)$
 and its time derivative $\dot{\Lambda}_1(t)$ for a time interval more or less corresponding
 to the bouncing phase. The time axis is the same as in the plots on the right in figure \ref{fig:background}.
 As explained in the caption of figure \ref{fig:background}, $\Lambda_2(t)$ is taken according to equation \eqref{eq:Lambda2ansatz} with
 $\Lambda_{2,0}=1.0\times 10^{-4}\,M_\mathrm{Pl}^4$, $\sigma=18.441\,t_\mathrm{Pl}$, and $t_B=0$.
 Then, the evolution of $\Lambda_1(t)$ and its time derivative follows from solving the Friedmann equations and the constraint
 equation \eqref{hcon2}.}
\label{fig:Lambdas}
\end{figure}

The above solutions can be verified numerically.
Setting the initial conditions in the second regime with a non-vanishing but initially sub-dominant cosmological constant $\Lambda_{1,0}$
and taking the ansatz for $\Lambda_2$ according to equation \eqref{eq:Lambda2ansatz},
we can numerically integrate the Friedmann equations \eqref{eq:Friedmann12} and \eqref{eq:Friedmann22}
together with the constraint equation \eqref{hcon2}.
The resulting dynamics is shown in figures \ref{fig:background} and \ref{fig:Lambdas}.
The background evolution clearly exhibits the phases of the matter bounce scenario with a non-singular bounce.
There is a first phase when $H<0$ and $\dot{H}<0$ where the universe is matter dominated and contracting
(see the plots on the left in figure \ref{fig:background} for $t\lesssim -40\,t_\mathrm{Pl}$).
Then around $t\approx -40\,t_\mathrm{Pl}$, $\Lambda_2$ becomes non-negligible (see figure \ref{fig:Lambdas}), and it triggers the NEC violating phase,
i.e.~the bouncing phase, during which $\dot H>0$ and $H$ transitions from being negative to positive
(see the plots on the right in figure \ref{fig:background}). After the bounce, the universe is expanding as in standard $\Lambda$CDM cosmology
since $\Lambda_2\rightarrow 0$ and $\dot{\Lambda}_1\rightarrow 0$ (see figure \ref{fig:Lambdas}),
i.e., the action reduces to standard Einstein gravity with a small positive constant.

\subsection{Reconstructing a potential for the time-like St\"{u}ckelberg scalar field}\label{sec:recovpot}

We saw in section \ref{sec:recovdiff} that we can associate $\Lambda_2=M^2\dot\varphi^0$ and $\Lambda_1=V(\varphi^0)$
together with the constraint $3H=V'(\varphi^0)/M^2$. With the ansatz of equation \eqref{eq:Lambda2ansatz} for $\Lambda_2(t)$,
this suggests $M^4\dot{f}(t)=\Lambda_{2,0}e^{-(t-t_B)^2/\sigma^2}$ [recall that at the background level $\varphi^0=M^2f(t)$].
After integration, this implies
\begin{equation}
\label{eq:fbackground}
 f(t)=f_i+\frac{\sqrt{\pi}}{2}\frac{\sigma\Lambda_{2,0}}{M^4}\bigg[\mathrm{erf}\left(\frac{t-t_B}{\sigma}\right)
 +\mathrm{erf}\left(\frac{t_B-t_i}{\sigma}\right)\bigg]\,,
\end{equation}
where we let $f_i\equiv f(t_i)$.
This is a monotonic function of $t$ as wanted.
Then, in the limit where $|t-t_B|\gg\sigma$, one finds that $f(t)\simeq\mathrm{constant}$,
and so, almost any potential $V(\varphi^0)$ will lead to $\Lambda_1\simeq\mathrm{constant}$ in that limit.
Thus, there is no important constraint on the potential from that regime.
Near the bounce, i.e.~for $|t-t_B|\ll\sigma$, equation \eqref{eq:fbackground} gives
\begin{equation}
 f(t)\simeq\tilde{f}_i+\frac{\Lambda_{2,0}}{M^4}(t-t_B)\,,
\end{equation}
where we let $\tilde{f}_i\equiv f_i+(\sqrt{\pi}/2)(\sigma\Lambda_{2,0}/M^4)\mathrm{erf}[(t_B-t_i)/\sigma]$.
Then, a good ansatz for the potential is
\begin{equation}
 V(\varphi^0)=A+\frac{1}{2}m_{\varphi^0}^2\left(\varphi^0\right)^2\,.
\end{equation}
Indeed, this implies
\begin{align}
 \Lambda_1(t)&=V(M^2f(t)) \nonumber \\
 &\simeq A+\frac{1}{2}m_{\varphi^0}^2\left(M^2\tilde{f}_i+\frac{\Lambda_{2,0}}{M^2}(t-t_B)\right)^2\,.
\end{align}
In comparison with equation \eqref{eq:Lambda1bounce}, which is an expression for $\Lambda_1(t)$ in the same limit,
we see that we must set $A=C=-\rho_0(a_0/a_B)^3$, $m_{\varphi^0}^2\Lambda_{2,0}^2/(2M^4)=3\Lambda_{2,0}^2/(4M_\mathrm{Pl}^2)$,
and $\tilde{f}_i=0$.
This last condition is equivalent to demanding $f_i=-(\sqrt{\pi}/2)(\sigma\Lambda_{2,0}/M^4)\mathrm{erf}[(t_B-t_i)/\sigma]$,
which is to say that we set the integration constant in equation \eqref{eq:fbackground} such that $f(t_B)=0$.
In sum, a good potential is
\begin{equation}
\label{eq:Vcov}
 V(\varphi^0)=\frac{1}{2}m_{\varphi^0}^2\left(\varphi^0\right)^2-\rho_0\left(\frac{a_0}{a_B}\right)^3\,,
\end{equation}
where the `mass' of the time-like St\"{u}ckelberg scalar field is $m_{\varphi^0}=\sqrt{3/2}M^2/M_\mathrm{Pl}$; and the time evolution
of the field $\varphi^0=M^2f(t)$ is given by
\begin{equation}
\label{eq:fbackgroundf}
 f(t)=\frac{\sqrt{\pi}}{2}\frac{\sigma\Lambda_{2,0}}{M^4}\mathrm{erf}\left(\frac{t-t_B}{\sigma}\right)\,.
\end{equation}
We can check that equation \eqref{eq:Vcov} implies $V'(\varphi^0)=3M^4\varphi^0/(2M_\mathrm{Pl}^2)$, and so,
the constraint equation $3H=V'(\varphi^0)/M^2$ gives
\begin{equation}
 3H(t)=\frac{3M^4f(t)}{2M_\mathrm{Pl}^2}\simeq\frac{3\Lambda_{2,0}}{2M_\mathrm{Pl}^2}(t-t_B)
\end{equation}
in the limit where $|t-t_B|\ll\sigma$. This is in agreement with equation \eqref{eq:Hbounce}.
Let us also note that by combining equations \eqref{eq:Vcov} and \eqref{eq:fbackgroundf} we find
\begin{align}
 \Lambda_1(t)&=V(M^2f(t)) \nonumber \\
 &=\frac{3\pi\sigma^2\Lambda_{2,0}^2}{16M_\mathrm{Pl}^2}\mathrm{erf}^2\left(\frac{t-t_B}{\sigma}\right)
 -\rho_0\left(\frac{a_0}{a_B}\right)^3\,.
\end{align}
Thus, in the limit where $|t-t_B|\gg\sigma$, one finds $\Lambda_1\simeq 3\pi\sigma^2\Lambda_{2,0}^2/(16M_\mathrm{Pl}^2)-\rho_0(a_0/a_B)^3$,
which can be a small positive constant if the parameters are tuned appropriately.
This is what we see in figure \ref{fig:Lambdas} where the shape of $\Lambda_1$ resembles an error function squared,
and $\Lambda_1$ is asymptotically a positive but small constant.

\section{Cosmological perturbation analysis}\label{sec:pertanalysis}

We now consider the linear cosmological perturbations of the theory about an FLRW background.
Due to the $SO(3)$ rotational symmetry of the background spacetime, one can decompose the metric perturbations
into scalar, vector, and tensor modes, and the helicities completely decouple at the linear perturbation level.
We define the metric perturbations as follows,
\begin{align}
 g_{00}=&-\left(1+2\alpha\right)\,, \nonumber \\
 g_{0i}=&~a\left(S_i+\partial_i\beta\right)\,, \nonumber \\
 g_{ij}=&~a^2\Big(\delta_{ij}+2\psi\delta_{ij}+\partial_i\partial_jE+\frac{1}{2}(\partial_iF_j+\partial_jF_i)+\gamma_{ij}\Big)\,,
\label{eq:pertmetric}
\end{align}
where $\alpha$, $\beta$, $\psi$ and $E$ are scalar perturbations, $S_i$ and $F_i$ are vector perturbations,
and the $\gamma_{ij}$'s are tensor perturbations. Vector modes satisfy the transverse conditions,
\begin{equation}
\partial_iS^i=\partial_iF^i=0\,,
\end{equation}
and tensor modes satisfy the transverse and traceless conditions, 
\begin{equation}
\label{eq:transversetraceless}
 \partial_i\gamma^{ij}=0\,, \qquad \gamma_i{}^i=0\,.
\end{equation}
Since we work in unitary gauge, the perturbations of all four St\"{u}ckelberg scalar fields are turned off. 

As we will see, the perturbation analysis reveals that there are only 2 tensor modes in the gravity sector, and
there is no scalar and vector graviton. This is consistent with the Hamiltonian analysis of section \ref{sec:Hamiltoniananalysis}.

\subsection{Scalar perturbations}\label{sec:scalarpert}

The derivation of the second-order perturbed action for scalar modes can be found in appendix \ref{sec:derivS2S}.
Consistent with the Hamiltonian analysis in section \ref{sec:Hamiltoniananalysis}, no helicity-$0$ mode of the graviton is spotted
in our perturbative expansion. The only scalar perturbation in our theory is the one from the matter sector,
which is represented by a canonical scalar field. The resulting perturbed action in Fourier space is
\begin{equation}
\label{eq:S2curvpert}
 S^{(2)}_\mathrm{scalar}=\int\mathrm{d}^3\mathbf{k}\,\mathrm{d}t~\left(\mathcal{K}\dot{\mathcal{R}}_k^2-\tilde{\Omega}\mathcal{R}_k^2\right)\,,
\end{equation}
where $\mathcal{R}_k$ is the Fourier transform of the curvature perturbation [defined in equation \eqref{eq:defcurvaturepert}]
with wavenumber $k$. The expressions for $\mathcal{K}$ and $\tilde{\Omega}$ are given in equations \eqref{eq:calK} and \eqref{eq:Omegatilde2}.

During a matter contracting phase, if we assume that $\Lambda_2\ll M_\mathrm{Pl}^2\dot{H}$ and thus
$M_\mathrm{Pl}^2\dot{H}\simeq -\frac{1}{2}\dot{\phi}^2$, the quadratic action of scalar perturbations simplifies to 
\begin{equation}
 S^{(2)}_\mathrm{scalar}\simeq M_\mathrm{Pl}^2\int\mathrm{d}^3\mathbf{k}\,\mathrm{d}t~a^3\epsilon\left(\dot{\mathcal{R}}_k^2-\frac{k^2}{a^2}\mathcal{R}_k^2\right)\,,
\end{equation}
where $\epsilon=\dot{\phi}^2/(2M_\mathrm{Pl}^2H^2)$.
This is the standard perturbed action for curvature perturbations in general relativity with a canonical scalar field
(see, e.g., \cite{Brandenberger:2003vk,Baumann:2009ds}).
In other words, the scalar perturbation is the same as the one in general relativity as long as
$\Lambda_1, \Lambda_2\ll M_\mathrm{Pl}^2H^2$ during the matter contracting phase,
no matter how large the graviton mass $m_g^2$ is.

During a bouncing phase, one would need to know the exact time dependence of $\mathcal{K}$ and $\tilde\Omega$ in order to solve
the EOM that results from varying the action of equation \eqref{eq:S2curvpert}. However, since the curvature perturbations of observational
interest remain mostly on super-Hubble scales during a non-singular bouncing phase,
they tend to remain constant \cite{nonsingularbouncepertanalysis,Quintin:2015rta},
i.e., their amplitude and spectral shape are unaffected. One can understand this fact by realizing that the duration of the bouncing phase
is usually much shorter than the wavelength of the perturbations that are considered, hence they cannot receive significant amplification.
In fact, it has been shown that the curvature perturbations can grow at most linearly in time\footnote{One caveat,
though, is the possible presence of gradient instabilities (see, e.g., \cite{Koehn:2015vvy}),
in which case modes with shorter wavelength would grow exponentially with time with an
exponent which increases as $k$ increases.
In the model presented here, gradient instabilities are expected to be absent from the EFT point of view.
From that perspective, the graviton is a gauge boson which couples to the density fluctuations from the matter sector
(in fact, the matter fluctuation could be considered as a Nambu-Goldston boson), and the coupling would occur
at the mixing scale corresponding to $\Lambda_\mathrm{mix}^2\sim \dot{H}$.
These two sectors are well decoupled on scales well above the mixing scale $\Lambda_\mathrm{mix}$.
Indeed, an explicit computation tells us that in the limit where $(k/a)^2\gg\dot{H}$,
we have a free field theory at leading order with positive sound speed equal to unity during the bouncing phase,
hence the theory is free from the possibly pathological
ultraviolet (UV) gradient instabilities.},
and the amplification that can be received is therefore bounded from above by the duration of the bouncing phase \cite{Quintin:2015rta}.
Consequently, we do not perform the full analysis of the evolution of curvature perturbations during
the non-singular bouncing phase is what follows and only assume that they remain unchanged through the bounce.

\subsection{Vector perturbations}\label{sec:vecpert}

Using the same methodology as the one described in appendix \ref{sec:derivS2S} for scalar modes,
one finds that the quadratic action of vector perturbations reads (again in momentum space, but we omit the subscript $k$ to simplify
the notation)
\begin{align}\label{actv2}
 S^{(2)}_\mathrm{vector}=&~\frac{1}{4}M_\mathrm{Pl}^2\int\mathrm{d}^3\mathbf{k}\,\mathrm{d}t~\left(\frac{1}{4}k^2a^3\dot{F}_i\dot{F}^i
 -k^2a^2S_i\dot{F}^i\right. \nonumber \\
 &\left.+\,k^2aS_iS^i-\frac{1}{4}m_g^2a^3F_iF^i\right)\,.
\end{align}
The vector mode $S_i$ does not have a kinetic term, and it gives us the constraint equation
\begin{equation}
 S_i=\frac{1}{2}a\dot{F}_i\,.
\end{equation}
Substituting the above solution back into the action of equation \eqref{actv2}, one finds
\begin{equation}
 S^{(2)}_\mathrm{vector}=-\frac{1}{16}M_\mathrm{Pl}^2\int\mathrm{d}^3\mathbf{k}\,\mathrm{d}t~m_g^2 k^2a^3 F_iF^i\,.
\end{equation}
This clearly shows that the kinetic term for vector perturbations has canceled out. It is by no means an accident, because a kinetic term of
vector modes is prohibited by the internal symmetry $\varphi^a\to\varphi^a+\Xi^a(\varphi^0)$ [recall equation \eqref{eq:sym1}].
After integrating out $F_i$, the whole action for vector perturbations vanishes, and there is no vector mode left in the theory. 
This result is also consistent with the Hamiltonian analysis of section \ref{sec:Hamiltoniananalysis}.

\subsection{Tensor perturbations}

Following the methodology described in appendix \ref{sec:derivS2S} but for tensor modes,
one finds that the perturbed action for tensor perturbations reads\footnote{In this sub-section,
we perform the analysis starting with the gravity action of equation \eqref{eq:actionmoregeneral},
which allows for a non-trivial propagation speed of gravitational waves.}
\begin{equation}
 S^{(2)}_\mathrm{tensor}=\frac{M_\mathrm{Pl}^2}{8}\int\mathrm{d}^3\mathbf{x}\,\mathrm{d}t~a^3
 \Big[c_1\dot{\gamma}_{ij}^2-c_2\frac{\left(\partial_l\gamma_{ij}\right)^2}{{a^2}}-m_g^2\gamma_{ij}^2\Big]\,.
\end{equation}
Converting to conformal time $\tau$ defined by $\mathrm{d}\tau\equiv a^{-1}\mathrm{d}t$, we can write
\begin{equation}
 S^{(2)}_\mathrm{tensor}=\frac{M_\mathrm{Pl}^2}{8}\int\mathrm{d}^3\mathbf{x}\,\mathrm{d}\tau~a^2
 \Big[c_1\gamma_{ij}'^2-c_2(\partial_l\gamma_{ij})^2-m_g^2a^2\gamma_{ij}^2\Big]\,.
\end{equation}
Letting
\begin{equation}
 \gamma_{ij}(\tau,\mathbf{x})=\int\frac{\mathrm{d}^3\mathbf{k}}{(2\pi)^{3/2}}\sum_{\lambda=+,\times}\epsilon_{ij}^{(\lambda)}
 \gamma_{\mathbf{k},\lambda}(\tau)e^{i\mathbf{k}\cdot\mathbf{x}}\,,
\end{equation}
where $\epsilon_i^{(\lambda)i}=0$, $k^i\epsilon_{ij}^{(\lambda)}=0$, and $\epsilon^{(\lambda)}_{ij}\epsilon^{(\lambda')ij}=2\delta_{\lambda\lambda'}$, the action becomes
\begin{align}
 S^{(2)}_\mathrm{tensor}=&~\frac{M_\mathrm{Pl}^2}{4}\sum_\lambda\int\mathrm{d}^3\mathbf{k}\,\mathrm{d}\tau~a^2
 \Big[c_1(\gamma_{\mathbf{k},\lambda}')^2 \nonumber \\
 &-\left(c_2k^2+m_g^2a^2\right)(\gamma_{\mathbf{k},\lambda})^2\Big]\,,
\end{align}
where $\gamma_{\mathbf{k},\lambda}$ represents the two polarization states of the tensor modes, the $+$ and $\times$ polarizations.
Varying the above action and defining $c_g^2\equiv c_2/c_1$, one obtains the EOM
\begin{equation}
 \gamma_k''+2\frac{a'}{a}\gamma_k'+\frac{c_1'}{c_1}\gamma_k'+c_g^2k^2\gamma_k+\frac{1}{c_1}m_g^2a^2\gamma_k=0
\end{equation}
for each polarization state $\lambda$.
From here on, let us assume that $c_1$ is a constant, and let us normalize it to unity, so $c_1\equiv 1$.
Therefore,
\begin{equation}
\label{eq:gammakEOM}
 \gamma_k''+2\frac{a'}{a}\gamma_k'+c_g^2k^2\gamma_k+m_g^2a^2\gamma_k=0\,,
\end{equation}
where $c_g(\tau)=\sqrt{c_2(\tau)}$.
Then, defining the Mukhanov-Sasaki variable
\begin{equation}
\label{eq:defMSu}
 u_{\mathbf{k},\lambda}\equiv\frac{M_\mathrm{Pl}}{2}a\gamma_{\mathbf{k},\lambda}\,,
\end{equation}
we can rewrite the action as
\begin{align}
 S^{(2)}_\mathrm{tensor}=&~\frac{1}{2}\sum_\lambda\int\mathrm{d}^3\mathbf{k}\,\mathrm{d}\tau~\Big[(u_{\mathbf{k},\lambda}')^2 \nonumber \\
 &-\Big(c_g^2k^2+m_g^2a^2-\frac{a''}{a}\Big)(u_{\mathbf{k},\lambda})^2\Big]\,.
\end{align}
Upon variation, the EOM becomes
\begin{equation}
\label{eq:ukgen}
 u_k''+\left(c_g^2k^2+m_g^2a^2-\frac{a''}{a}\right)u_k=0\,,
\end{equation}
again for each polarization state $\lambda$. We note that the differential equation is of the form $u_k''+\omega_k^2u_k=0$ with
effective time-dependent frequency given by
\begin{equation}
\label{eq:efffreqgen}
 \omega_k^2(\tau)=c_g^2k^2+m_g^2a^2-\frac{a''}{a}\,.
\end{equation}
In comparison with the tensor mode EOM in general relativity \cite{Brandenberger:2003vk},
we notice that our theory allows for a non-trivial speed of sound for the gravitational waves,
but most importantly, the effective frequency picks up a mass term.
Therefore, while the mass term in the action of the form of equation \eqref{eq:Lmassterm}
did not affect the background dynamics and did not introduce additional DOFs for scalar, vector, and tensor perturbations,
we notice that it can affect the evolution of gravitational waves.
This will be crucial in the context of the matter bounce scenario as we will soon see in section \ref{sec:evogw}.

\section{Anisotropies}\label{sec:anisotropies}

While the previous section explored the characteristics of inhomogeneities about an FLRW background,
one may also consider the theory in an anisotropic background.
Let us slightly deform the FLRW metric by introducing anisotropies as follows,
\begin{equation}
\label{eq:metricani}
 g_{\mu\nu}\mathrm{d}x^\mu\mathrm{d}x^\nu=-\mathrm{d}t^2+a(t)^2\sum_{i=1}^3e^{2\theta_i(t)}\mathrm{d}x^i\mathrm{d}x^i\,,
\end{equation}
with the constraint equation $\sum_i\theta_i=0$. This represents a metric of Bianchi type-I form.
One can think of $a(t)$ as the average scale factor
of the universe, and the $\theta_i(t)$'s are corrections (or anisotropies) to the average expansion or contraction.
Accordingly, $H=\dot a/a$ is the mean Hubble parameter, and the Hubble parameter along a given spatial direction $x^i$
is \cite{Cai:2013vm} $H_i=H+\dot{\theta}_i$.

Linearly expanding our action of equation \eqref{mmg1} with the metric of equation \eqref{eq:metricani},
where we think of the $\theta_i$'s as small anisotropies about an FLRW background, we find
\begin{equation}
\label{eq:Sani}
 S_\theta\supset M_\mathrm{Pl}^2\int\mathrm{d}^3\mathbf{x}\,\mathrm{d}t~a^3\left(\dot{\theta}_i^2-m_g^2\theta_i^2\right)\,,
\end{equation}
where we see that the anisotropies receive a mass term (just like tensor modes). The corresponding EOM reads
\begin{equation}
\label{eq:sigmagen}
 \ddot{\theta}_i+3H\dot{\theta}_i+m_g^2\theta_i=0\,,
\end{equation}
and if one associates an energy-momentum tensor with the variation of $S_\theta$, then the anisotropies carry an energy density given by
\begin{equation}
\label{eq:rhoanigen}
 \rho_\theta=\frac{M_\mathrm{Pl}^2}{2}\sum_{i=1}^3\left(\dot\theta_i^2+m_g^2\theta_i^2\right)\,.
\end{equation}
It makes sense that the anisotropies pick up a mass term just like tensor modes because the two are very similar.
Indeed, one can think of the anisotropies as traceless ($\sum_i\theta_i=0$) and transverse ($\partial_i\theta^i=0$)
just like tensor modes [c.f.~equation \eqref{eq:transversetraceless}]. Accordingly, taking the limit $k\rightarrow 0$
of equation \eqref{eq:gammakEOM} for tensor modes and converting from conformal time to physical time,
one finds the same EOM as equation \eqref{eq:sigmagen} for anisotropies.

In comparison with the case where the graviton is massless, the metric of equation \eqref{eq:metricani} in general relativity yields
the Lagrangian density $\mathcal{L}_\theta\supset M_\mathrm{Pl}^2\dot\theta_i^2$ and the EOM $\ddot\theta_i+3H\dot\theta_i=0$,
which usually suggests that $\dot\theta_i\propto a^{-3}$, hence $\rho_\theta\sim M_\mathrm{Pl}^2\dot\theta_i^2\propto a^{-6}$
(see, e.g., \cite{Cai:2013vm}).
This is reminiscent of a massless canonical scalar field, e.g., call it $\Theta$ with Lagrangian $\mathcal{L}=(1/2)(\partial_\mu\Theta)^2$,
which at the background level has an effective EoS $p_\Theta=\rho_\Theta=(1/2)\dot\Theta^2$
(in agreement with the energy density scaling as $a^{-6}$).
With the massive gravity action of equation \eqref{mmg1}, a comparison with equation \eqref{eq:Sani} tells us that the
anisotropies behave like a massive scalar field
with Lagrangian $\mathcal{L}=(1/2)(\partial_\mu\Theta)^2-(1/2)m_\Theta^2\Theta^2$,
where the `mass' of the anisotropies is the mass of the graviton, i.e.~$m_\Theta=m_g$.
In that case, the energy density of the massive scalar field is
$\rho_\Theta=(1/2)\dot\Theta^2+(1/2)m_\Theta^2\Theta^2$ [in agreement with equation \eqref{eq:rhoanigen}],
and so the energy density no longer necessarily grows as $a^{-6}$ in a contracting universe.
In fact, if $m_\Theta\gg|H|$, we expect to recover the result of a coherently oscillating massive scalar field in cosmology,
in which case the energy density would scale as $a^{-3}$. In other words, the anisotropies would behave as pressureless matter.
This will be shown explicitly in the next section.

\section{Solutions in a matter-dominated contracting phase}\label{sec:solmatter}

We derived the general EOMs for tensor perturbations and anisotropies in the previous sections
for our massive gravity theory. We now want to solve these equations during a matter-dominated contracting phase,
so let us set up the background evolution during such a phase. The scale factor is given by $a(t)\propto(-t)^{2/3}$ and
so the Hubble parameter is
\begin{equation}
 H(t)=\frac{2}{3(t-\tilde{t}_{B-})}
\end{equation}
for $t<\tilde{t}_{B-}$. Here, $\tilde{t}_{B-}$ denotes the time at which the singularity would be reached if no new physics appeared at
high energy scales to violate the NEC and avoid a Big Crunch. Without loss of generality, we set $\tilde{t}_{B-}=0$ in what follows.

\subsection{Evolution of anisotropies}\label{sec:evoani}

Let us first solve the EOM for anisotropies [equation \eqref{eq:sigmagen}] in a matter-dominated contracting phase.
Since $H=2/(3t)$, the differential equation is
\begin{equation}
 \ddot{\theta_i}+\frac{2}{t}\dot{\theta_i}+m_g^2\theta_i=0\,,
\end{equation}
whose general solution is
\begin{equation}
 \theta_i(t)=\frac{1}{(-t)}\left\{C_1\cos[m_g(-t)]+C_2\sin[m_g(-t)]\right\}\,,
\end{equation}
where we have assume that the graviton mass $m_g^2$ is a constant.
For simplicity, let us consider one of two modes only, e.g., $\theta_i(t)=C_1\cos(m_g t)/t$
(we would find the same result below if we considered the other mode or both).
Then, according to equation \eqref{eq:rhoanigen}, the energy density for the anisotropy $\theta_i$ is given by
\begin{align}
 \rho_{\theta_i}=&~\frac{M_\mathrm{Pl}^2}{2}\left(\dot\theta_i^2+m_g^2\theta_i^2\right) \nonumber \\
 =&~\frac{M_\mathrm{Pl}^2m_g^2}{2}\left(\frac{C_1}{-t}\right)^2\Bigg[1+\frac{3}{2}\left(\frac{H}{m_g}\right)\sin[2m_g(-t)] \nonumber \\
 &+\frac{9}{4}\left(\frac{H}{m_g}\right)^2\cos^2[m_g(-t)]\Bigg]\,,
\end{align}
where we used the fact that $H=2/(3t)$. If we consider the limit where $m_g\gg|H|$, then it follows that
\begin{equation}
\label{eq:rhothetaiam3}
 \rho_{\theta_i}\simeq\frac{M_\mathrm{Pl}^2m_g^2}{2}\left(\frac{C_1}{-t}\right)^2\propto a^{-3}
\end{equation}
since $a(t)\propto(-t)^{2/3}$.
Therefore, if the mass of the graviton $m_g$ is larger than the absolute value of the Hubble parameter in
some time interval in a matter-dominated contracting universe, then we find that the total energy density in anisotropies
$\rho_\theta=\sum_i\rho_{\theta_i}$ scales as $a^{-3}$, i.e., it grows at the same rate as the background energy density of
the pressureless matter.

The above result implies that as long as the anisotropies are sub-dominant at some initial time in the far past,
they will always remain sub-dominant (in the regime when $m_g\gg|H|$).
Furthermore, one can even show that this result is independent of the background EoS.
Indeed, as long as $m_g\gg|H|$, one finds that $\rho_\theta\propto a^{-3}$ for any background EoS parameter
$w\equiv p/\rho\geq 0$. The proof can be found in appendix \ref{sec:rhoanievogen}.
For example, if the model included radiation in addition to pressureless matter, radiation would get to dominate at higher
energy scales, and the already sub-dominant anisotropies would be washed out since
the energy density of radiation ($\rho_\mathrm{rad}\propto a^{-4}$) would grow faster than that of anisotropies
($\rho_\theta\propto a^{-3}$). As a result, the model is free of the BKL instability.

We note that the above resolution to the BKL instability problem in a contracting universe is also free of the problems
that one encounters in trying to resolve the BKL instability with an Ekpyrotic phase of contraction as in
\cite{matterEkpyrotic,Cai:2013vm}. In that context, the Ekpyrotic scalar field has an EoS parameter $w\gg 1$,
and thus, the background energy density in an Ekpyrotic phase of contraction scales as $a^{-3(1+w)}$,
i.e., it grows much faster than the energy density in anisotropies that scales as $a^{-6}$ when the graviton is massless.
While this appropriately washes out the anisotropies for an Ekpyrotic field with isotropic pressure \cite{Erickson:2003zm,Cai:2013vm},
it has been shown in \cite{Barrow:2015wfa} that the presence of dominant ultra-stiff pressure anisotropies,
which one should certainly expect in an anisotropic background, does not necessarily lead to isotropization on approach to a bounce.
Indeed, the presence of anisotropic pressures in the background fluid would add a source term on the right-hand side of
equation \eqref{eq:sigmagen}, which could enhance the growth of anisotropies.
It would be interesting to reproduce the analysis of \cite{Barrow:2015wfa} with our massive gravity theory,
but at least in the context of a matter-dominated contracting universe, one would not expect important anisotropic pressures
to source the right-hand side of equation \eqref{eq:sigmagen} since a pressureless fluid does not have any pressure by definition.
Therefore, our resolution of the BKL instability with massive gravity should be robust with regards to that issue
in the matter bounce scenario.

Resolving the BKL instability with an Ekpyrotic phase of contraction after a matter-dominated contracting phase also introduces a
fine-tuning problem. As explained in \cite{Levy:2016xcl}, in order to generate $N$ e-folds of matter contraction with sub-dominant anisotropies,
one needs the initial ratio of the energy density in anisotropies to that in matter to be smaller than $e^{-6N}$, which can be an extremely small
number. With our massive gravity theory, we showed that the energy densities in anisotropies and matter grow at the same rate,
so their ratio remains constant. Thus, one only has to require the initial ratio to be smaller than unity, which is a much smaller fine-tuning
requirement compared to a factor of order $e^{-6N}$.

\subsection{Evolution of gravitational waves}\label{sec:evogw}

To solve for the evolution of tensor modes in a matter-dominated contracting phase, we need to solve equation \eqref{eq:ukgen},
which we recall is of the form $u_k''+\omega_k^2u_k=0$ with effective frequency given by equation \eqref{eq:efffreqgen}.
Converting the scale factor evolution $a(t)\propto(-t)^{2/3}$ to conformal time with $\mathrm{d}\tau=a^{-1}\mathrm{d}t$, one finds
\begin{equation}
 a(\tau)=\left(\frac{-\tau}{\tau_m}\right)^2\,,\qquad \tau<0\,,
\end{equation}
where $\tau_m$ is some constant at which $a(\tau_m)=1$ by convention. Then,
\begin{equation}
\label{eq:confH}
 \mathcal{H}\equiv\frac{a'}{a}=\frac{2}{\tau}
\end{equation}
is the conformal Hubble parameter. It is also useful to calculate
\begin{equation}
\label{eq:appoa}
 \frac{a''}{a}=\frac{2}{\tau^2}\,.
\end{equation}
Noting that $\mathcal{H}=aH$ and using equations \eqref{eq:confH} and \eqref{eq:appoa}, we note that
\begin{equation}
 \frac{1}{a^2}\frac{a''}{a}=\frac{H^2}{\mathcal{H}^2}\frac{2}{\tau^2}=\frac{H^2}{2}\,,
\end{equation}
and thus, the physical effective frequency of equation \eqref{eq:efffreqgen} can be written as
\begin{equation}
\label{eq:omegak2oagen}
 \frac{\omega_k^2}{a^2}=c_g^2\frac{k^2}{a^2}+m_g^2-\frac{1}{a^2}\frac{a''}{a}=c_g^2\frac{k^2}{a^2}+m_g^2-\frac{H^2}{2}\,.
\end{equation}
This suggests that the size of the graviton mass compared to the Hubble parameter (in absolute value) will determine the evolutionary
behavior of the tensor modes. Thus, we separate the analysis depending on whether the graviton mass is small or large compared to $|H|$.

\subsubsection{Small graviton mass regime}\label{sec:evogwsmallmg}

In this subsection, we will assume that $m_g\ll|H|$ within some time interval during which the nearly scale-invariant power spectrum
of curvature perturbations is generated, i.e.~during which the scales of observational interest exit the Hubble radius.

Equation \eqref{eq:omegak2oagen} suggests that we can write the effective frequency as
\begin{equation}
 \omega_k^2=c_g^2k^2-\left(\frac{1}{2}-\frac{m_g^2}{H^2}\right)\mathcal{H}^2\,,
\end{equation}
using the property $\mathcal{H}=aH$ once again. Recalling equation \eqref{eq:confH}, we can finally write
\begin{equation}
 \omega_k^2=c_g^2k^2-\frac{\nu_t^2-\frac{1}{4}}{\tau^2}\,,
\end{equation}
where we defined the index
\begin{equation}
\label{eq:defnut}
 \nu_t\equiv\sqrt{\frac{9}{4}-4\frac{m_g^2}{H^2}}\,.
\end{equation}
Provided $m_g\ll|H|$, we can assume that $\nu_t$ is approximately constant (and not complex) in the time interval of interest.
Let us further assume that $c_g$ is a constant, i.e.~$c_g'(\tau)=0$. In fact, we will most often set $c_g=1$,
but we do the calculation in more generality.
Therefore, the solutions to the EOM,
\begin{equation}
 u_k''+\left(c_g^2k^2-\frac{\nu_t^2-\frac{1}{4}}{\tau^2}\right)u_k=0\,,
\end{equation}
are the Hankel (or, equivalently, Bessel) functions of the first and second kind:
\begin{equation}
 u_k(\tau)=\sqrt{-\tau}\left\{A_kH_{\nu_t}^{(1)}[c_gk(-\tau)]+B_kH_{\nu_t}^{(2)}[c_gk(-\tau)]\right\}\,.
\end{equation}
In the limit where $c_gk|\tau|\gg 1$, we impose Bunch-Davies initial conditions,
$u_k(\tau)\simeq e^{-ic_gk\tau}/\sqrt{2c_gk}$,
which sets the integration constants as follows: $A_k=(\sqrt{\pi}/2)e^{i\vartheta}$
and $B_k=0$, with $\vartheta\equiv (\pi/2)({\nu_t}-3/2)$.
Therefore, the above solution becomes
\begin{equation}
 u_k(\tau)=\frac{\sqrt{\pi}}{2}e^{i\vartheta}\sqrt{-\tau}H_{\nu_t}^{(1)}[c_gk(-\tau)]\,.
\end{equation}
Expanding on large scales, where $c_gk|\tau|\ll 1$, and keeping only the growing mode of the Hankel function that scales as $(-\tau)^{-{\nu_t}}$
(there is also a decaying mode that scales as $(-\tau)^{{\nu_t}}$; recall here that $-\tau\rightarrow 0^+$ in a contracting universe,
and from equation \eqref{eq:defnut}, ${\nu_t}>0$), one finds
\begin{equation}
\label{eq:uksolsmallmg}
 u_k(\tau)\simeq -\frac{i2^{{\nu_t}-1}e^{i\vartheta}\Gamma({\nu_t})}{\sqrt{\pi}}c_g^{-{\nu_t}}k^{-{\nu_t}}(-\tau)^{1/2-{\nu_t}}\,,
\end{equation}
where $\Gamma$ denotes the gamma function.
Then, the corresponding power spectrum of tensor perturbations defined on large scales is
\begin{equation}
\label{eq:defPt}
 \mathcal{P}_{t}(k,\tau)=2\times\mathcal{P}_\gamma(k,\tau)=2\times\left(\frac{2}{aM_\mathrm{Pl}}\right)^2\times\mathcal{P}_u(k,\tau)\,,
\end{equation}
where
\begin{equation}
\label{eq:defPu}
 \mathcal{P}_u(k,\tau)\equiv\frac{k^3}{2\pi^2}\left|u_k(\tau)\right|^2\,,
\end{equation}
and $u_k(\tau)$ is given by equation \eqref{eq:uksolsmallmg}.
The first factor of $2$ in equation \eqref{eq:defPt} accounts for the two polarization states of the gravitational waves,
and the factor of $2/(aM_\mathrm{Pl})$ is due to the conversion from the Mukhanov-Sasaki variable
to the variable $\gamma$ [recall equation \eqref{eq:defMSu}].
The resulting power spectrum of tensor perturbations on large scales is
\begin{equation}
 \mathcal{P}_{t}(k,\tau)=\frac{4^{{\nu_t}}\Gamma({\nu_t})^2}{\pi^3c_g^{2{\nu_t}}}\frac{k^{3-2{\nu_t}}(-\tau)^{1-2{\nu_t}}}{a^2M_\mathrm{Pl}^2}\,.
\end{equation}

Already, we can calculate the tensor spectral tilt:
\begin{equation}
 n_t\equiv\frac{\mathrm{d}\ln\mathcal{P}_t}{\mathrm{d}\ln k}=3-2{\nu_t}=3-2\sqrt{\frac{9}{4}-4\frac{m_g^2}{H^2}}\,.
\end{equation}
Since we are working in the limit where $m_g\ll|H|$, the above can be simplified to
\begin{equation}
\label{eq:ntsmallmg}
 n_t\simeq\frac{8}{3}\frac{m_g^2}{H^2}\,.
\end{equation}
It is thus clear that the tensor tilt is small and positive, i.e., it is a blue tilt (opposite to the tilt
one obtains in inflationary cosmology \cite{Baumann:2009ds}, but the same sign as the tilt obtained in string
gas cosmology \cite{Brandenberger:2006xi}).
Also, in the limit where $m_g\rightarrow 0$, we recover $n_t\rightarrow 0$, i.e.~a scale-invariant power spectrum, as expected.

Using the property $\mathcal{H}=aH$ and equation \eqref{eq:confH}, one can re-express the amplitude of the power spectrum as
\begin{equation}
\label{eq:Ptfinalsmallmg}
 \mathcal{P}_t(k,t)=\frac{2\,\Gamma({\nu_t})^2}{\pi^3c_g^{2{\nu_t}}}\left(\frac{k}{aM_\mathrm{Pl}}\right)^{3-2{\nu_t}}
 \left(\frac{|H(t)|}{M_\mathrm{Pl}}\right)^{2{\nu_t}-1}\,.
\end{equation}
We note that $k$ is the comoving wavenumber here, so $k/a$ represents a physical wavenumber.
In the limit where $m_g\rightarrow 0$, we have ${\nu_t}\rightarrow 3/2$, and at the time $t_{B-}$ at the end of the contracting
phase (or right before the bouncing phase), one finds
\begin{equation}
\label{eq:Ptfinalzeromg}
 \mathcal{P}_t(k,t_{B-})\rightarrow\frac{1}{2\pi^2c_g^3}\left(\frac{H_B}{M_\mathrm{Pl}}\right)^2\,,
\end{equation}
where we defined $H_B\equiv H(t_{B-})$, i.e., it is the energy scale of the bounce.
This matches the usual result (see, e.g., \cite{Li:2016xjb}) with $c_g=1$.
We evaluate the power spectrum at the end of the contracting phase because this is the point at which the amplitude stops increasing.
As we argued in section \ref{sec:scalarpert}, the perturbations will remain more or less constant during the non-singular bouncing phase,
so we can immediately match the above power spectrum with the primordial power spectrum after the bounce, i.e., at the beginning of
standard Big Bang cosmology. Thus, we will drop the argument $t_{B-}$ when it is clear that we are talking about the primordial
power spectrum that can be connected with observations.

In comparison to the tensor power spectrum,
the primordial power spectrum of curvature perturbations generated during a matter-dominated contracting phase on super-Hubble scales
is given by (see, e.g., \cite{Li:2016xjb})
\begin{equation}
\label{eq:PRmb}
 \mathcal{P}_\mathcal{R}(k)=\frac{1}{48\pi^2c_s}\left(\frac{H_B}{M_\mathrm{Pl}}\right)^2\,.
\end{equation}
In order to avoid the over-production of scalar non-Gaussianities in the matter-dominated contracting phase, it has been shown
that $c_s$ cannot be too small. In fact, in order to satisfy the current constraints on $f_\mathrm{NL}$, one must have \cite{Li:2016xjb}
$c_s\gtrsim 0.62$. Imposing $\mathcal{P}_\mathcal{R}(k)=A_s\approx 2.2\times 10^{-9}$ (to match the amplitude observed by
Planck \cite{Ade:2015xua}), one would therefore need $|H_B|\gtrsim 8.0\times 10^{-4}\,M_\mathrm{Pl}$.
Comparing equations \eqref{eq:Ptfinalsmallmg} and \eqref{eq:PRmb}, the tensor-to-scalar ratio is given by
\begin{align}
 r_\star&\equiv \frac{\mathcal{P}_t(k_\star)}{\mathcal{P}_\mathcal{R}(k_\star)} \nonumber \\
 &=\frac{96\,\Gamma({\nu_t})^2}{\pi}\frac{c_s}{c_g^{2{\nu_t}}}\left(\frac{k_\star}{M_\mathrm{Pl}}\right)^{3-2{\nu_t}}
 \left(\frac{|H_B|}{M_\mathrm{Pl}}\right)^{2{\nu_t}-3}\,,
\end{align}
where $k_\star$ represents the physical wavenumber at which the observation is made.
Let us set $c_g=1$, $c_s=0.62$, and $|H_B|=8.0\times 10^{-4}\,M_\mathrm{Pl}$ to have an idea of the size of the
above tensor-to-scalar ratio.
Furthermore, let us say that we consider a mode with physical wavenumber $k_\star$ that exits the Hubble radius at a time $t_\star$
at which point $m_g/|H|=0.13$. Then, ${\nu_t}\approx 1.477$, and the above expression for $r_\star$ becomes
\begin{equation}
 r_\star\approx 20.546\left(\frac{k_\star}{M_\mathrm{Pl}}\right)^{0.0454}\approx 0.048\left(\frac{k_\star}{0.05\,\mathrm{Mpc}^{-1}}\right)^{0.0454}\,.
\end{equation}
Thus, we see that for modes of observational interest, the tensor-to-scalar ratio is suppressed within the current
observational bound (for $k_\star=0.05\,\mathrm{Mpc}^{-1}$, $r_\star<0.07$ at 95\% confidence \cite{Array:2015xqh}).
However, this is true only if $m_g$ is sufficiently non-zero, because
as $m_g\rightarrow 0$, one finds ${\nu_t}\rightarrow 3/2$ and $r\rightarrow 24c_s/c_g^3$
on all scales and independently of the value of $H_B$. With the constraint $c_s\gtrsim 0.62$, this cannot satisfy current observational
bounds without tuning $c_g\gtrsim 5.968$, and such a large super-luminal propagation speed of gravitational waves does not sound
very realistic.

\subsubsection{Large graviton mass regime}\label{sec:evogwlargemg}

Let us now explore the possibility that $m_g\gg|H|$, and as before, we assume that this is valid
within some time interval during which the nearly scale-invariant power spectrum
of curvature perturbations is generated. Then, there are three regimes to be considered (we assume that $c_g=1$ from here on):
\begin{enumerate}
 \item $(k/a)^2\gg m_g^2\gg H^2\implies \omega_k^2\simeq k^2$;
 \item $m_g^2\gg (k/a)^2\gg H^2\implies \omega_k^2\simeq m_g^2a^2$;
 \item $m_g^2\gg H^2\gg (k/a)^2\implies \omega_k^2\simeq m_g^2a^2$.
\end{enumerate}
Regimes (1) and (2) represent sub-Hubble modes and regime (3) represents super-Hubble modes in the conventional sense.
However, the scale of interest here is $m_g$. It separates the evolution of the perturbations into actually only two regimes.
There are the `sub-graviton' scales, where $(k/a)^2\gg m_g^2$ (this is regime (1) above), in which case the modes are deeply sub-Hubble;
and there are the `super-graviton' scales, where $(k/a)^2\ll m_g^2$.

On sub-graviton scales, the EOM for tensor modes reads
$u_k''+k^2u_k=0$, so the general solution is $u_k(\tau)=C_{1,k}e^{-ik\tau}+C_{2,k}e^{ik\tau}$.
We require the usual Bunch-Davies normalization, which sets the integration constants $C_{1,k}=(2k)^{-1/2}$ and $C_{2,k}=0$, so that
\begin{equation}
\label{eq:BD}
 u_k(\tau)=\frac{1}{\sqrt{2k}}e^{-ik\tau}\,.
\end{equation}
On super-graviton scales, the EOM reads $u_k''+m_g^2a^2u_k=0$, hence the differential equation becomes
\begin{equation}
 u_k''+m_g^2\left(\frac{\tau}{\tau_m}\right)^4u_k=0\,,
\end{equation}
and the solutions are Bessel functions,
\begin{align}
 u_k(\tau)=&~\left(\frac{m_g(-\tau)^3}{6\tau_m^2}\right)^{1/6}\left[A_kJ_{-1/6}\left(\frac{m_g(-\tau)^3}{3\tau_m^2}\right)\right. \nonumber \\
 &\left.+B_kJ_{1/6}\left(\frac{m_g(-\tau)^3}{3\tau_m^2}\right)\right]\,,
\label{eq:uksollargemggen}
\end{align}
where $A_k$ and $B_k$ are two integration constants.
One can show that $m_g(-\tau)^3/(3\tau_m^2)=(2/3)(m_g/|H|)$, so in the limit where $m_g\gg|H|$, the argument of the Bessel functions is large
compared to unity. Thus, the Bessel functions can be expanded to find
\begin{align}
 u_k(\tau)\simeq&~\frac{1}{\sqrt{\pi}}\left(\frac{6\tau_m^2}{m_g(-\tau)^3}\right)^{1/3}\left[A_k\cos\left(\frac{\pi}{6}
 -\frac{m_g(-\tau)^3}{3\tau_m^2}\right)\right. \nonumber \\
 &\left.+B_k\sin\left(\frac{\pi}{6}+\frac{m_g(-\tau)^3}{3\tau_m^2}\right)\right]
\label{eq:solsuper}
\end{align}
in the limit where $m_g\gg|H|$.

We now match the solutions to the EOM on sub- and super-graviton regimes at ``graviton horizon crossing'',
i.e.~when the perturbation wavelength is equal to the graviton Compton wavelength.
This is the case when $k/a=m_g$, which happens at the time $(-\tau)=\tau_m\sqrt{k/m_g}$ for a given mode with comoving wavenumber $k$.
Specifically, we equate equation \eqref{eq:solsuper} with the Bunch-Davies vacuum equation \eqref{eq:BD}
at the time of graviton horizon crossing,
and we do the same thing with their conformal time derivatives.
One obtains a set of two equations that can be solved for the unknowns $A_k$ and $B_k$ to find
\begin{align}
 A_k&=\frac{\sqrt{2\pi}\tau_m^{1/3}}{6^{1/3}m_g^{1/6}}\exp\left[\frac{i}{3}\left(4\vartheta_k+\frac{\pi}{2}\right)\right]\,, \\
 B_k&=-\frac{\sqrt{2\pi}\tau_m^{1/3}}{6^{1/3}m_g^{1/6}}\exp\left[\frac{i}{3}\left(4\vartheta_k+\pi\right)\right]\,,
\end{align}
where we defined $\vartheta_k\equiv k^{3/2}\tau_m/m_g^{1/2}$.
Therefore, the solution to the EOM on super-graviton scales, which reduces to the properly normalized Bunch-Davies vacuum on sub-graviton scales,
is given by
\begin{align}
 u_k(\tau)\simeq&~\sqrt{\frac{2}{m_g}}\frac{\tau_m}{(-\tau)}e^{4i\vartheta_k/3}
 \left[e^{i\pi/6}\cos\left(\frac{\pi}{6}-\frac{m_g(-\tau)^3}{3\tau_m^2}\right)\right. \nonumber \\
 &\left.-e^{i\pi/3}\sin\left(\frac{\pi}{6}+\frac{m_g(-\tau)^3}{3\tau_m^2}\right)\right]\,,
\end{align}
and what is physically relevant for the power spectrum is the modulus squared:
\begin{equation}
 |u_k(\tau)|^2\simeq\frac{2}{m_g}\left(\frac{\tau_m}{-\tau}\right)^2\left(\frac{1}{4}\right)=\frac{1}{2m_ga(\tau)}\,.
\end{equation}
Finally, the power spectrum is [recall the definition from equations \eqref{eq:defPt} and \eqref{eq:defPu}]
\begin{equation}
\label{eq:Ptlargemg}
 \mathcal{P}_t(k,t)=\frac{8}{a^2M_\mathrm{Pl}^2}\frac{k^3}{2\pi^2}\frac{1}{2m_ga(t)}
 =\frac{2}{\pi^2}\frac{(k/a)^3}{M_\mathrm{Pl}^2m_g}\,,
\end{equation}
which is a highly blue spectrum (the tilt is $n_t=3$).
Dividing by equation \eqref{eq:PRmb}, we get the tensor-to-scalar ratio:
\begin{equation}
 r_\star=96c_s\frac{k_\star^3}{m_gH_B^2}\,.
\end{equation}
For instance, if we set $c_s=0.62$ and $|H_B|=8.0\times 10^{-4}\,M_\mathrm{Pl}$, we have
\begin{equation}
 r_\star\approx 2\times 10^{-163}\left(\frac{k_\star}{0.05\,\mathrm{Mpc}^{-1}}\right)^3\left(\frac{10^{-3}\,M_\mathrm{Pl}}{m_g}\right)\,.
\end{equation}
Therefore, with a typical pivot scale $k_\star=0.05\,\mathrm{Mpc}^{-1}$
[and more generally for physically observable scales in the cosmic microwave background (CMB)]
and with a large graviton mass of the order of $|H_B|$,
the tensor-to-scalar ratio is highly suppressed, well below current observational bounds.
The model would effectively predict no observable primordial B-mode polarization in the CMB, similar to the prediction in
Ekpyrotic cosmology \cite{Lehners:2008vx} and pre-Big Bang cosmology \cite{preBigBang}.

\subsubsection{Connection with observations}

Let us now connect the above results with cosmological observations. Currently, we note that
the most constraining bound on the graviton mass is\footnote{Latest bounds from gravitational waves observations
\cite{LIGObounds,Baker:2017hug} are of the same order.} \cite{deRham:2016nuf}
\begin{equation}
\label{eq:mgconstraint}
 m_g<7.2\times10^{-23}\,\mathrm{eV}\ (95\%\ \mathrm{C.\,L.})\,.
\end{equation}
Let us first consider the possibility that $m_g=\mathrm{constant}$ throughout comic history, including the matter-dominated contracting phase
in the context of the matter bounce scenario. Let us say that $m_g=7.0\times 10^{-23}\,\mathrm{eV}$, i.e.~right below current constraints.
Then, this means that for $|H(t)|<7.0\times 10^{-32}\,\mathrm{GeV}$, the mass of the graviton is effectively large compared to the background
Hubble parameter in absolute value, and so, the primordial power spectrum of gravitational waves is given by equation \eqref{eq:Ptlargemg},
which we can rewrite as follows,
\begin{equation}
 \mathcal{P}_t(k_\mathrm{p})\approx 1.269\times 10^{-127}\left(\frac{k_\mathrm{p}}{0.05\,\mathrm{Mpc}^{-1}}\right)^3\,,
\end{equation}
where $k_\mathrm{p}\equiv k/a$ represents the physical wavenumber at the time of Hubble radius crossing.
The above thus applies only for modes with $k_\mathrm{p}<m_g=7.0\times 10^{-32}\,\mathrm{GeV}\approx 5.489\times 10^7\,\mathrm{Mpc}^{-1}$.
Therefore, this applies for really all of the observables modes in the CMB (the CMB includes modes in the approximate range
$[10^{-4},10^0]\,\mathrm{Mpc}^{-1}$). Therefore, we see from the above that the primordial gravitational wave spectrum is highly
suppressed: even for the observable modes on the smallest CMB length scales, e.g., $k_\mathrm{p}\sim 1\,\mathrm{Mpc}^{-1}$,
we have $\mathcal{P}_t\sim 10^{-123}$.
It is only in the far UV, for $k_\mathrm{p}>m_p$, that the blue spectrum becomes closer to scale invariant (but still with a blue tilt).
This corresponds to the regime where modes exit the Hubble radius when the graviton mass is effectively small compared
to the background Hubble parameter in absolute value, so the primordial gravitational wave power spectrum
has a blue tilt given by equation \eqref{eq:ntsmallmg}.
In fact, the power spectrum is asymptotically scale invariant for $k_\mathrm{p}\rightarrow\infty$
[recall equation \eqref{eq:Ptfinalzeromg}].

In the previous setup though, i.e.~with a constant graviton mass that satisfies equation \eqref{eq:mgconstraint},
the anisotropies grow in a controlled way only for $|H(t)|\ll m_g$, but for $|H(t)|\gtrsim m_g$
they can rapidly dominate over the background and lead to the known BKL instability before the bounce.
Indeed, we recall that the results of section \ref{sec:evoani}, in particular equation \eqref{eq:rhothetaiam3}, are only valid if $m_g\gg|H|$.
Therefore, it would be preferable if one had $m_g\gg |H(t)|$ for the whole contracting phase, i.e.~for $0<|H(t)|\leq |H_B|$.
In that case, it is natural to take $m_g>|H_B|$.
As a result, the anisotropies are under control for the whole contracting phase (all the way to the bounce),
and the gravitational wave power spectrum is given by equation \eqref{eq:Ptlargemg}, i.e., it is suppressed
and scales as $k_\mathrm{p}^3$ across all scales (for $0<k_\mathrm{p}<|H_B|$).
However, since we expect $|H_B|$ to be relatively large, $m_g$ cannot be constant throughout cosmic history in
that case; $m_g$ would have to be a time-dependent function in the EFT.
For example, if we take $|H_B|\sim 8\times 10^{-4}\,M_\mathrm{Pl}$, then we could have $m_g=m_{g,0}\sim 10^{-3}\,M_\mathrm{Pl}$
for the whole contracting phase and $m_g$ would have to go to zero [or at least below the constraint of equation \eqref{eq:mgconstraint}]
rapidly before or after the bounce\footnote{Most current constraints on the graviton mass apply only today in cosmic history,
but a non-trivial graviton mass across time in standard Big Bang cosmology would still leave observable imprints in different CMB observations
(see, e.g., \cite{Brax:2017pzt}). Therefore, the graviton mass must already be sufficiently small after the bounce.}.

The action of equation \eqref{mmg1} allows for a time-dependent function $m_g(t,Z^{ab},\delta_{ab})$.
However, the analyses in sections \ref{sec:evoani}, \ref{sec:evogwsmallmg},
and \ref{sec:evogwlargemg} assumed that $m_g$ was constant in time,
so we must set the functional form of $m_g(t,Z^{ab},\delta_{ab})$ so that $m_g$ is approximately constant (and large) before the bounce, quickly transitions
at the beginning of the bounce and becomes very small (or zero) during the bouncing phase and for the rest of cosmic evolution.
For instance, we could have
\begin{equation}
 m_g(t,Z^{ab},\delta_{ab})=\frac{m_{g,0}}{2}\left[1-\mathrm{erf}\left(\frac{t-t_B}{\sigma}\right)\right]\,.
\end{equation}
in which case for $|t-t_B|\gg\sigma$ and $t<t_B$ (i.e.~before the bounce),
we get $m_g(t)\simeq m_{g,0}$,
and similarly, for $|t-t_B|\gg\sigma$ and $t>t_B$ (i.e.~after the bounce), we get $m_g(t)\simeq 0$.
The transition time between the two constant mass phases would be of the order of $\sigma$, i.e.~of the order of the duration of the bouncing phase.

One can then recover the space-time diffeomorphism invariance of the action by constructing the appropriate function
$F(\varphi^0,Z^{ab},\delta_{ab})$ so that equation \eqref{eq:mgtcov} matches the above functional form of $m_g(t,Z^{ab},\delta_{ab})$.
It is straightforward to reconstruct such a function with a similar analysis to the one performed in section \ref{sec:recovpot},
although the resulting form of $F$ might be complicated due to the appearance of the error function.
However, we recall that the Gaussian ansatz in equation \eqref{eq:Lambda2ansatz} as well as the subsequent error functions
were taken for simplicity to make the different limits explicit. One could very well reconstruct the diffeomorphism-invariant action
with different choices of functions such as rational functions that still yield the desired background dynamics.

\section{Conclusions and discussion}\label{sec:conclusions}

We introduced a particular modified gravity model with a massive graviton,
and we showed that it only propagates two gravitational degrees of freedom and is free of ghosts.
We also showed that this model admits homogeneous and isotropic cosmological solutions with a non-singular bounce.
We further studied the evolution of cosmological perturbations in this model. Whereas the scalar
cosmological fluctuations grow on super-Hubble scales in the contracting phase like in
other models, the finite graviton mass suppresses the growth of the gravitational waves.
The finite graviton mass also enters in the EOM for anisotropies and leads
to the conclusion that the energy density in anisotropies scales like pressureless matter.
Hence, our model admits a realization of the {\it matter bounce} scenario which is free
of two of the key problems of such scenarios: the \textit{large $r$ problem}
and the anisotropy problem (also known as the BKL instability).
Indeed, our model predicts a naturally highly suppressed tensor-to-scalar ratio on observational scales,
and there is no anisotropy problem (no BKL instability).

While the analysis performed in this paper assumed a particular model of massive gravity,
there is no \textit{a priori} reason why the main conclusions should not hold with a different theory of gravity,
as long as it admits a non-trivial mass for the graviton. For instance, the theories of \cite{dRGT,bigravity,Langlois:2014jba}
could all represent massive gravity theories in which the matter bounce scenario could be embedded to solve
the large $r$ and anisotropy problems. However, due to the existence of scalar and vector polarizations of the
graviton in these theories, the scale invariance and near Gaussianity
of the primordial curvature perturbations might be spoiled by the additional
DOFs in the gravity sector. It would be interesting to investigate these theoretical possibilities.

In this paper, we did not study the evolution of the cosmological perturbations through the non-singular bouncing phase,
although we argued that they should remain more or less unchanged. Yet, a proper analysis should be done in a follow-up paper.
Accordingly, one may also wish to properly compute the strong energy scale of the theory, which determines the range of validity of the EFT.
This is of particular interest when studying perturbations in non-singular bouncing cosmology in the context of EFT
(see, e.g., \cite{Koehn:2015vvy,deRham:2017aoj}).
Finally, it may also be interesting to extend the analysis of this paper to explore non-Gaussianities from primordial perturbations
generated during the matter contracting phase. We expect that at the three-point function level, similar to the case studied
in \cite{Domenech:2017kno}, the graviton mass term will only contribute to the scalar-scalar-tensor and scalar-tensor-tensor couplings.
However, due to the highly suppressed tensor perturbations, we do not expect sizable three-point correlation functions arising from those
graviton mass terms. Thus, the amplitude of non-Gaussianity at the three-point function level should be the same as the one in the
literature \cite{Cai:2009fn}, i.e.~$f_\mathrm{NL}\sim\mathcal{O}(1)$. Nevertheless, at the four-point function level,
there could be interesting and sizable observational effects due to the non-trivial graviton mass.
It would be interesting to investigate these non-linear effects and find new distinguishable features for very early universe models.
\newline

\begin{acknowledgments}
We thank Guillem Dom\`enech, Ryo Namba, Shi Pi, and Daisuke Yoshida for useful discussions,
and we wish to acknowledge the stimulating atmosphere at the COSMO-17 conference in Paris where this project was initiated.
CL is supported by the Japanese JSPS fellowship for overseas researchers, by the JSPS Grant-in-Aid for Scientific Research No.~15F15321,
and by POLONEZ 3 from the Polish National Science Centre.
JQ acknowledges financial support from the Vanier Canada Graduate Scholarship administered by
the Natural Sciences and Engineering Research Council of Canada (NSERC).
The research of RB is supported by an NSERC Discovery Grant and funds from the Canada Research Chair program.
\end{acknowledgments}

\onecolumngrid

\appendix

\section{Second-order perturbed action for scalar modes}\label{sec:derivS2S}

We start with the full action given by equation \eqref{mmg1} plus the matter action of equation \eqref{eq:Smatterfield},
i.e., we parametrize the matter content by a canonical scalar field $\phi$ with potential $U(\phi)$ for simplicity.
Then, we linearly perturb the metric as in equation \eqref{eq:pertmetric} and consider only the scalar perturbations.
Similarly, we linearly perturb the scalar field as
\begin{equation}
 \phi(t,\mathbf{x})=\bar{\phi}(t)+\delta\phi(t,\mathbf{x})\,.
\end{equation}
In what follows, we drop the overbar on $\bar\phi(t)$ when it is clear that we are only referring the background evolution of $\phi$.
Then, the second-order perturbed action of scalar perturbations can be decomposed as
\begin{equation}
 S^{(2)}_\mathrm{scalar}=S^{(2)}_\mathrm{gravity}+S^{(2)}_\mathrm{matter}\,,
\end{equation}
with
\begin{align}\label{sact2}
 S^{(2)}_\mathrm{gravity}=&~M_\mathrm{Pl}^2\int\mathrm{d}^3\mathbf{k}\,\mathrm{d}t~\left[-\frac{1}{12}m_g^2k^4E^2-3a^3\dot{\psi}^2+ak^2\psi^2
 +a^3k^2\dot{E}\dot{\psi}-\frac{1}{2}a^3Hk^2\psi\dot{E}-2a^2k^2\beta\dot{\psi}-3a^3H^2\alpha^2\right. \nonumber \\
 &\left.+\,\alpha\left(2a^2Hk^2\beta+2ak^2\psi-a^3Hk^2\dot{E}+6a^3H\dot{\psi}\right)
 -\frac{1}{2}E\left(3a^3H^2k^2\psi+a^3Hk^2\dot{\psi}\right)-\frac{1}{2}a^3\dot{H}k^2E\psi\right]\,, \\
 S^{(2)}_\mathrm{matter}=&~\int\mathrm{d}^3\mathbf{k}\,\mathrm{d}t~\left\{\frac{1}{2}a^3\delta\dot{\phi}^2
 -\frac{1}{2}a^3\dot{\phi}\left(k^2E+2\alpha-6\psi\right)\delta\dot{\phi}
 +\frac{1}{2}a^3\dot{\phi}^2\alpha^2-\frac{1}{2}a\delta\phi^2\left[k^2+a^2U''(\phi)\right]\right. \nonumber \\
 &\left.-\frac{1}{2}\delta\phi\left[2a^2\dot{\phi}k^2\beta+\left(k^2E-2\alpha-6\psi\right)\partial_t\left(a^3\dot{\phi}\right)\right]\right\}\,,
\end{align}
where we recall that $2M_\mathrm{Pl}^2\dot{H}=-\dot{\phi}^2+\Lambda_2$.
We note that the above action has been transformed to Fourier space where $\mathbf{k}$ represents the wavevector with magnitude $k\equiv|\mathbf{k}|$,
also known as the wavenumber,
and each perturbation variable represents its own Fourier transform, i.e., we omit the subscript $k$ on each perturbation variable
to simplify the notation in this appendix. Varying $S^{(2)}_\mathrm{scalar}$ with respect to $\alpha$ and $\beta$, one obtains
\begin{align}
\label{eq:a}
 \alpha=&~\frac{\dot{\psi}}{H}+\frac{\dot{\phi}\delta\phi}{2M_\mathrm{Pl}^2H}\,, \\
 \beta=&-\frac{\psi}{aH}+\frac{1}{2}a\dot{E}+\frac{a\dot{\phi}\delta\dot{\phi}}{2k^2M_\mathrm{Pl}^2H}
 -\frac{a\dot{\phi}^3\delta\phi}{4k^2M_\mathrm{Pl}^4H^2}
 -\frac{a\dot{\phi}^2\dot{\psi}}{2k^2M_\mathrm{Pl}^2H^2}-\frac{a\ddot{\phi}{\delta\phi}}{2k^2M_\mathrm{Pl}^2H}\,,
\label{eq:b}
\end{align}
which represent the Hamiltonian and momentum constraints, respectively.
Substituting equations \eqref{eq:a} and \eqref{eq:b} back into the quadratic action of scalar perturbations, one finds
\begin{align}
 S^{(2)}_\mathrm{scalar}=&~\int\mathrm{d}^3\mathbf{k}\,\mathrm{d}t~\left\{-\frac{1}{12}M_\mathrm{Pl}^2m_g^2k^4E^2
 +M_\mathrm{Pl}^2\frac{k^2a\dot{H}}{H^2}\psi^2
 +\frac{1}{2}a^3\delta\dot{\phi}^2+\frac{a^3\dot{\phi}^2}{2H^2}\dot{\psi}^2
 +a^3\delta\dot{\phi}\left(3\psi\dot{\phi}-\frac{\dot{\phi}\dot{\psi}}{H}\right)\right. \nonumber\\
 &+\frac{1}{2}a^3\delta\phi^2\left[-\frac{k^2}{a^3}+\frac{\dot{\phi}^4}{4M_\mathrm{Pl}^4H^2}
 +\frac{(-\dot{H}+6H^2)\dot{\phi}^2+4H\ddot{\phi}\dot{\phi}}{2M_\mathrm{Pl}^2H^2}+\frac{3\dot{H}\dot{\phi}
 +3H\ddot{\phi}+\dddot{\phi}}{\dot{\phi}}\right] \nonumber\\
 &\left.+\,a^3\delta\phi\left[\psi\left(\frac{k^2\dot{\phi}}{Ha^2}+9H\dot{\phi}+3\ddot{\phi}\right)
 +\dot{\psi}\left(\frac{\dot{\phi}^3}{2M_\mathrm{Pl}^2H^2}+\frac{3H\dot{\phi}+\ddot{\phi}}{H}\right)\right]\right\}\,.
\end{align}
We find that $E$ does not have a kinetic term, and thus, it yields the constraint $E=0$. Then, we introduce the curvature perturbation variable,
\begin{equation}
\label{eq:defcurvaturepert}
 \mathcal{R}\equiv \psi-\frac{H\delta\phi}{\dot{\phi}}\,,
\end{equation}
so the above action can be rewritten as
\begin{align}
 S^{(2)}_\mathrm{scalar}=\int\mathrm{d}^3\mathbf{k}\,\mathrm{d}t~&\left\{\frac{a^3\dot{\phi}^2}{2H^2}\dot{\mathcal{R}}^2+\Omega\mathcal{R}^2
 +\psi\left[\frac{a^3\Lambda_2\dot{\phi}^2}{2M_\mathrm{Pl}^2H^3}\dot{\mathcal{R}}
 -\frac{a^3\Lambda_2\dot{\phi}^2(6M_\mathrm{Pl}^2H^2+\Lambda_2)}{4M_\mathrm{Pl}^4H^4}\mathcal{R}\right]\right. \nonumber \\
 &\left.+\,\psi^2\left(\frac{a^3\Lambda_2^2\dot{\phi}^2}{8M_\mathrm{Pl}^4H^4}+\frac{ak^2\Lambda_2}{2H^2}
 +\frac{3a\Lambda_2a^2\dot{\phi}^2}{4M_\mathrm{Pl}^2H^2}\right)\right\}\,,
\label{act2h}
\end{align}
where $\Omega$ is a function of the background quantities $\dot{\phi}$, $\Lambda_2$, $H$, etc.
Although the above action depends on both $\mathcal{R}$ and $\psi$, we note that there is only one scalar DOF in the theory,
i.e.~the one from matter fluctuations. The graviton itself does not have a helicity-$0$ component. This is consistent with the
Hamiltonian analysis in section \ref{sec:Hamiltoniananalysis}.
The presence of $\psi$, which does not appear in general relativity, is a consequence of the broken
temporal diffeomorphism of the theory.
The variable $\psi$ is actually a Lagrangian multiplier, and it gives us the following constraint:
\begin{equation}
 \psi=\frac{\dot{\phi}^2\left(6M_\mathrm{Pl}^2H^2\mathcal{R}+\Lambda_2\mathcal{R}-2M_\mathrm{Pl}^2H\dot{\mathcal{R}}\right)}
 {4M_\mathrm{Pl}^4H^2\frac{k^2}{a^2}+6M_\mathrm{Pl}^2H^2\dot{\phi^2}+\Lambda_2\dot{\phi}^2}\,.
\end{equation}
Substituting the above constraint back into the perturbed action of equation \eqref{act2h}, one finds
\begin{equation}
 S^{(2)}_\mathrm{scalar}=\int\mathrm{d}^3\mathbf{k}\,\mathrm{d}t~\left(\mathcal{K}\dot{\mathcal{R}}^2-\tilde{\Omega}\mathcal{R}^2\right)\,,
\end{equation}
where $\mathcal{K}$ and $\tilde{\Omega}$ are given by
\begin{align}
\label{eq:calK}
 \mathcal{K}=&~\frac{2k^2M_\mathrm{Pl}^4a^3\dot{\phi}^2+3M_\mathrm{Pl}^2a^5\dot{\phi}^4}{4k^2M_\mathrm{Pl}^4H^2
 +a^2\left(6M_\mathrm{Pl}^2H^2+\Lambda_2\right)\dot{\phi}^2}\,, \\
 \tilde{\Omega}=&~\frac{k^2M_\mathrm{Pl}^2a^5\dot{\phi}^2}{\left(4k^2M_\mathrm{Pl}^4H^2+a^2\left(6M_\mathrm{Pl}^2H^2
 +\Lambda_2\right)\dot{\phi}^2\right)^2}\Bigg[\Lambda_2^2\left(\frac{4k^2M_\mathrm{Pl}^2}{a^2}+2\dot{\phi}^2\right)-M_\mathrm{Pl}^2H\dot{\Lambda}_2
 \left(\frac{4k^2M_\mathrm{Pl}^2}{a^2}+6\dot{\phi}^2\right) \nonumber \\
 &-M_\mathrm{Pl}^2H^2\Lambda_2\left(\frac{2k^2\dot{\phi}^2}{a^2H^2}+\left(\frac{8\ddot{\phi}}{H\dot{\phi}}+24\right)
 \frac{k^2M_\mathrm{Pl}^2}{a^2}+\frac{3\dot{\phi}^4}{M_\mathrm{Pl}^2H^2}+24\dot{\phi}^2\right)+2M_\mathrm{Pl}^2H^2
 \left(\frac{2k^2M_\mathrm{Pl}^2}{a^2}+3\dot{\phi}^2\right)^2\Bigg]\,,
\label{eq:Omegatilde2}
\end{align}
i.e., they are two functions of the background quantities $\dot{\phi}$, $\Lambda_2$, $H$, etc.
We immediately notice from the above that the scalar mode is always ghost free, because $\mathcal{K}>0$.
Also, in the UV limit where $k^2/a^2\gg\dot{H}$, we have
\begin{equation}
 S^{(2)}_\mathrm{scalar}\simeq M_\mathrm{Pl}^2\int\mathrm{d}^3\mathbf{k}\,\mathrm{d}t~a^3\epsilon\left(\dot{\mathcal{R}}^2-\frac{k^2}{a^2}\mathcal{R}^2\right)\,,
\end{equation}
where $\epsilon=\dot{\phi}^2/(2M_\mathrm{Pl}^2H^2)$, and thus, the theory is free from UV gradient instabilities.

\section{Evolution of anisotropies in a general background}\label{sec:rhoanievogen}

We consider a general contracting universe dominated by matter with EoS $p=w\rho$,
assuming $w\geq 0$. In this case, the scale factor and Hubble parameter are given by
\begin{equation}
\label{eq:generalaandH}
 a(t)\sim(-t)^{\frac{2}{3(1+w)}}\,,\qquad H(t)=\frac{2}{3(1+w)t}\,.
\end{equation}
Then, the general solution to equation \eqref{eq:sigmagen} is
\begin{equation}
\label{eq:thetagenw}
 \theta_i(t)=(-t)^{\nu_a}\left\{[C_1J_{\nu_a}[m_g(-t)]+C_2Y_{\nu_a}[m_g(-t)]\right\}\,,
\end{equation}
and its time derivative is
\begin{equation}
\label{eq:dotthetagenw}
 \dot\theta_i(t)=-m_g(-t)^{\nu_a}\left\{[C_1J_{\nu_a-1}[m_g(-t)]+C_2Y_{\nu_a-1}[m_g(-t)]\right\}\,,
\end{equation}
where $J_{\nu_a}$ and $Y_{\nu_a}$ are the Bessel functions of the first and second kind and where we defined the index
\begin{equation}
 \nu_a\equiv\frac{1}{2}-\frac{1}{1+w}\,.
\end{equation}
In the limit where $m_g\gg|H|$, we note that $m_g(-t)\gg 1$, and so, the asymptotic form of all the above Bessel functions
is $[m_g(-t)]^{-1/2}$, ignoring constant factors and oscillatory functions of time, i.e., we only stress the time dependence of the overall amplitude.
Thus, equations \eqref{eq:thetagenw} and \eqref{eq:dotthetagenw} give
\begin{equation}
 \theta_i(t)\sim m_g^{-1/2}(-t)^{\nu_a-1/2}\,, \qquad \dot\theta_i(t)\sim m_g^{1/2}(-t)^{\nu_a-1/2}\,,
\end{equation}
and so, the energy density given by equation \eqref{eq:rhoanigen} becomes
\begin{equation}
 \rho_\theta\sim\dot\theta_i^2+m_g^2\theta_i^2\sim m_g(-t)^{2\nu_a-1}\,,
\end{equation}
where the power of the time dependence can be expanded in terms of the EoS parameter $w$ as
\begin{equation}
 2\nu_a-1=-\frac{2}{1+w}\,.
\end{equation}
Using equation \eqref{eq:generalaandH} to convert the time dependence back into a scale factor dependence, one finds
\begin{equation}
 \rho_\theta\sim(-t)^{-\frac{2}{1+w}}\sim\left(a^{\frac{3(1+w)}{2}}\right)^{-\frac{2}{1+w}}=a^{-3}\,.
\end{equation}
Therefore, this shows that the energy density in anisotropies grows as $a^{-3}$ in a contracting universe no matter what is the
EoS of the dominant background matter content provided $m_g\gg|H|$.

\twocolumngrid

\end{document}